\documentclass{scrartcl}

\usepackage{array}
\usepackage{color}
\usepackage{colinmacros}

\newcommand{\aeff}{\alpha_{\mathrm{eff}}}

\usepackage{mathtools}
\usepackage{amsmath}
\usepackage{amssymb}
\usepackage{amsthm}
\usepackage{xcolor}
\input{tudcolours.def}
\usepackage[printonlyused,smaller]{acronym}

\makeatletter
\let\oldabs\abs
\def\abs{\@ifstar{\oldabs}{\oldabs*}}

\newcommand{\defeq}{\coloneqq}

\newcommand{\differential}[1]{\mathrm{d} #1}

\newcommand{\eqstop}{.}
\newcommand{\eqcomma}{,}

\newcommand{\ie}{\textit{i.e.}}
\newcommand{\eg}{\textit{e.g.}}

\usepackage{hyperref}
\usepackage{url}
\usepackage{cleveref}
\usepackage{tikz}
\usetikzlibrary{arrows,shapes,backgrounds}

\newtheorem{remark}{Remark}
\newtheorem{lemma}{Lemma}

\newtheorem{corollary}{Corollary}
\newtheorem{pseudocode}{Pseudocode}

\begin{document}
  \title{Likelihood-Free Dynamical Survival Analysis Applied to the COVID-19 Epidemic in Ohio}
  
  \author{Colin Klaus \href{https://orcid.org/0000-0001-9759-4068}{\includegraphics[width=3mm]{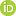}}, The Ohio State University\\ 
  Matthew Wascher \href{https://orcid.org/0000-0002-4948-7724}{\includegraphics[width=3mm]{ORCID-iD_icon-16x16.png}}, University of Dayton\\ 
  Wasiur R. KhudaBukhsh \href{https://orcid.org/0000-0003-1803-0470}{\includegraphics[width=3mm]{ORCID-iD_icon-16x16.png}}, The University of Nottingham\\ 
  Grzegorz A. Rempa{\l}a \href{https://orcid.org/0000-0002-6307-4555}{\includegraphics[width=3mm]{ORCID-iD_icon-16x16.png}}, The Ohio State University }
  
  \maketitle

  \begin{abstract}
  The \ac{DSA} is a framework for  modeling  epidemics based on mean field dynamics applied to individual (agent) level history of infection and recovery.  Recently, \ac{DSA} has been  shown  to be an effective tool in analyzing complex non-Markovian epidemic processes that are  otherwise difficult to handle using  standard methods.  One of the advantages of \ac{DSA} is its  representation of  typical epidemic data in a   simple although not explicit form that  involves solutions of certain differential equations. In this work we describe how a complex  non-Markovian \ac{DSA} model may be applied  to a specific data set with the help of appropriate numerical and statistical schemes. The ideas are illustrated with a data example of the COVID-19 epidemic in Ohio. 
  \\
  \end{abstract}

  \section{Introduction} 
  
   As of July, 2022, the coronavirus disease 2019 (COVID-19) pandemic, caused by the  severe acute respiratory syndrome coronavirus 2 (SARS-CoV-2), has taken more than six million lives worldwide. In response to the pandemic, scientists from all disciplines have made a concerted effort to address the ever growing analytic challenges of prediction, intervention, and control. The mathematical tools employed range from the purely deterministic \ac{ODE}/\ac{PDE} models to describe population dynamics (at the macro or ecological scale) to fully stochastic agent-based models (at the micro scale); from physics-inspired mechanistic models, both stochastic and deterministic, to purely statistical approaches such as ensemble models.  However, despite the longstanding history of the discipline of mathematical epidemiology and the enormous recent efforts, the pandemic has laid bare crucial gaps in the state-of-the-art methodology. 
   
   While the macro models are simple to interpret and easy to calibrate, the micro agent-based models provide more flexibility to model elaborate what-if scenarios. In a similar vein, the mechanistic models provide insights into the underlying biology and epidemiology of the disease but are often outperformed by purely statistical methods when it comes to accuracy in prediction and forecasting. While there is no obvious way to completely bypass such trade-offs between these often diametrically opposed modeling approaches, the \acf{DSA} method  \cite{choi2019modeling,KhudaBukhsh2020InterfaceFocus,DiLauro2022nonMarkov,Wascher2021IDSA,OSU_whitepaper,Harley2022Ebola,KhudaBukhsh2021Prison,Caleb2020Stone,KhudaBukhsh2022Israel}, a survival analytic statistical method derived from dynamical systems, holds some promise. The present paper is about a likelihood-free means of performing \ac{DSA} in the context of non-Markovian models of infectious disease epidemiology. More specifically, we develop a \ac{DSA} method for a non-Markovian epidemic model with vaccination based on the \ac{ABC} framework. 
   
   {Why non-Markov models?} The standard compartmental Markovian models, which employ \acp{CTMC} to keep track of counts of individuals in different compartments (\eg, individuals with different immunological statuses), assume the infectious period and the contact interval \cite{kenah2011biostat,DiLauro2022nonMarkov} are exponentially distributed and are thus characterized by a constant hazard function. This simplistic assumption almost always misrepresents the underlying biology of the disease and hence, is untenable. See \cite{DiLauro2022nonMarkov,vanKampen1998nonMarkov} for a detailed discussion on this point. Also, see \cite[Table 1 and Figure 1]{kenah2011biostat} for a numerical illustration of the bias in the estimates of model parameters if a Markovian model is wrongly assumed when the underlying model is non-Markovian. 
   
   A popular and analytically convenient approach to building more realistic non-Markovian models is to make use of the general theory of  measure-valued processes that keep track of not only the population counts but also additional covariates such as  individual's  age. Here  the word ``age'' is used as an  umbrella term  to refer to the physical age of the individual, the age of infection, time since vaccination etc. While the measure-valued processes do require more mathematical sophistication than their \ac{CTMC} counterparts, as we discuss in \Cref{sec:nonMarkovModel}, the age-stratified population densities can be described by a comparatively simple system of \acp{PDE} in the limit of a large population \cite{DiLauro2022nonMarkov,Franco2021Renewal,Hyman2007Infection_age,Sherborne2018nonMarkov}. The crux of the \ac{DSA} method is to interpret this  mean-field limiting system of \acp{PDE} as describing probability distributions of transfer times (the time required to move from one compartment to another). We shall make this notion clear in \Cref{subsec:DSA} and discuss the statistical benefits of the \ac{DSA} method in \Cref{sec:discussion}. 
   
   As an illustration of the \ac{ABC}-based \ac{DSA} method, we apply it to the COVID-19 epidemic in the state of Ohio, USA. The method is shown to fit the real case count data well and capture nontrivial trends. Detailed numerical results are provided in \Cref{sec:num_results}. In addition to the introduction of \ac{ABC}-\ac{DSA} methodology, our second contribution in this paper is a solution method for the mean-field limiting system of \acp{PDE} with nonlocal boundary conditions. For the sake of algorithmic implementation, we also present it in the form of a pseudocode. An implementation in the Julia programming language is  made available for the wider community.  

  The rest of the paper is structured as follows: The stochastic non-Markovian epidemic model is described in \Cref{sec:nonMarkovModel}. The mean-field limit in the form of a system of \acp{PDE} and the \ac{DSA} methodology are also described in \Cref{sec:nonMarkovModel}. In \Cref{sec:PDE_solution}, we describe the main technical details of how we solve the limiting mean-field \ac{PDE} system. We describe the statistical approach to parameter inference in \Cref{sec:prm}, followed by numerical results in \Cref{sec:num_results}. Finally, we conclude with a brief discussion in \Cref{sec:discussion}. For the sake of completeness, additional mathematical details, numerical results, and figures are provided in the Appendix.

  \section{Non-markovian mass-action model}\label{sec:nonMarkovModel}
  
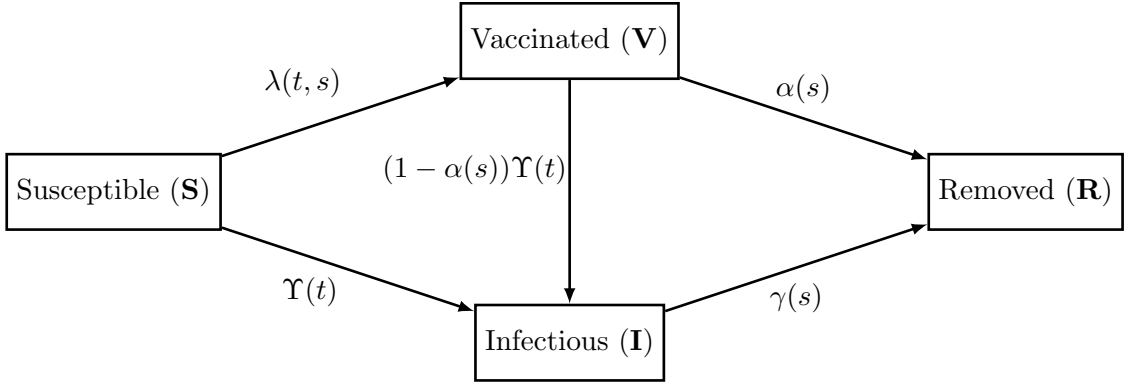
\begin{figure}
        \centering
        \begin{tikzpicture}[scale=1.0, every node/.style={transform shape}]
        \begin{scope}
            \node(box1)[draw,rectangle, minimum size=1cm,line width=1pt] at (0,0){Susceptible (\textbf{S})};
            \node(box2)[draw,rectangle, minimum size=1cm,line width=1pt] at (6,-2){Infectious (\textbf{I})};
            \node(box3)[draw,rectangle, minimum size=1cm, line width=1pt] at (12.0,0){Removed (\textbf{R})};
            \node(box4)[draw,rectangle, minimum size=1cm, line width=1pt] at (6,2){Vaccinated (\textbf{V})};

            \draw  [-latex, line width=1pt](box1) to node [minimum size=0.3cm, xshift=-0.5cm, yshift=-0.30cm] {$\Upsilon(t)$} (box2);
            \draw  [-latex, line width=1pt](box1) to node [minimum size=0.3cm, xshift=-0.5cm, yshift=0.450cm] {$\lambda(t,s)$} (box4);
            \draw  [-latex, line width=1pt](box2) to node [minimum size=0.3cm, xshift=0.0cm, yshift=-0.40cm] {$\gamma(s)$} (box3);
            \draw  [-latex, line width=1pt](box4) to node [minimum size=0.3cm, xshift=0.0cm, yshift=0.40cm] {$\alpha(s)$} (box3);
            \draw  [-latex, line width=1pt](box4) to node [minimum size=0.3cm, xshift=-1.25cm, yshift=0.30cm] {$(1-\alpha(s))\Upsilon(t)$} (box2);
            
            \end{scope}
        \end{tikzpicture}
        \caption{\label{fig:sir}%
          Diagrammatic representation of the non-Markovian compartmental model. At time $t$, a susceptible individual with age $s$ moves either to a vaccinated compartment   with instantaneous rate $\lambda(t,s)$, or to the infectious compartment with instantaneous rate $\Upsilon(t) \defeq \int_0^\infty\beta(s)y_{I}(t,s)\differential{s}$. The vaccinated individuals are also subject to infection. An infected individual at age $s$ can be removed with instantaneous rate $\gamma(s)$. The vaccine efficacy $\alpha$ is assumed age-dependent.
          }
      \end{figure}

  Our stochastic model is adapted from \cite{DiLauro2022nonMarkov}. As shown in \Cref{fig:sir}, the model has four compartments: \textbf{S}usceptible, \textbf{V}accinated, \textbf{I}nfectious, and \textbf{R}emoved. Individuals are in exactly one of the four compartments. Upon vaccination, susceptible individuals move to the \textbf{V} compartment. We assume only susceptible individuals are vaccinated. Both (unvaccinated) susceptible and vaccinated individuals can get infected, in which case they move to the \textbf{I} compartment. Finally, infected individuals either recover or are removed. In either case, they move to the terminal compartment \textbf{R}.  In addition to the counts of individuals in the four compartments, we keep track of the age distribution. Here, the word ``age'' refers to the physical age for the susceptible individuals, time since vaccination for the vaccinated individuals, time since infection or age of infection for the infected individuals, and finally, time since recovery or removal for the removed individuals. The age of the removed individuals are of no interest because the removed individuals do not contribute to the dynamics. It is possible to include other important covariates, such as sex, comorbidity, in the model. However, for the sake of simplicity, we only keep track of  age. Suppose we have $n$ individuals. 
  
  The age distribution of individuals in different compartments is  described in terms of finite, point measure-valued processes whose atoms are individual ages. See \cite{DiLauro2022nonMarkov}. Such an approach has been previously used to model age-stratified \ac{BD} processes \cite{Tran2008limit,Tran2009traits}, spatially stratified populations \cite{Fournier2004microscopic}, delays in \acp{CRN} \cite{KhudaBukhsh2020Delay}. The instantaneous rates of jump are assumed to depend on the individual ages. In particular, a susceptible individual of age $s$ has an instantaneous rate $\lambda(t,s)$ of getting vaccinated at time $t$. Moreover, a susceptible individual is also subject to an infection pressure exerted by the infectious individuals. We denote the hazard function for the probability distribution of the contact interval \cite{DiLauro2022nonMarkov,kenah2011biostat} by $\beta$. Therefore, an infectious individual of age of infection $s$ makes an infectious contact with a susceptible or vaccinated individual at an instantaneous rate $\beta(s)$. The hazard function of the probability law of the infectious period is denoted by $\gamma$. The vaccine efficacy is also assumed to depend on the age of the vaccinated individual (time since vaccination). At age $s$, a vaccinated individual moves directly to the \textbf{R} compartment at rate unity with probability $\alpha(s)$, while with probability $(1-\alpha(s))$, they are subject to the same infection pressure as the susceptible individuals and can get infected at the same instantaneous rate, which is the population sum total, scaled by $1/n$, of the $\beta$'s evaluated at the individual ages of infection. 
  
  In the limit of a large population ($n\to \infty$) and under suitable constraints on the hazard functions $\beta, \gamma, \lambda$, the measure-valued processes, when appropriately scaled by $n^{-1}$, can be shown to converge to deterministic continuous measure-valued functions \cite{DiLauro2022nonMarkov,Tran2008limit,Fournier2004microscopic,Tran2009traits,KhudaBukhsh2020Delay}. The densities (with respect to the Lebesgue measure) of those limiting measure-valued functions can be then described in terms of a system of \acp{PDE}.  Let us define
  \begin{itemize}
      \item $y_{S}(t,s)$: The density of susceptible individuals with age $s$ at time $t$;
      \item $y_{V}(t,s)$: The density of vaccinated individuals with age (of vaccination) or time since vaccination $s$ at time $t$;
      \item $y_{I}(t,s)$: The density of infectious individuals with age (of infection) or time since infection $s$ at time $t$;
      \item $y_{R}(t)$: The proportion of removed individuals. 
  \end{itemize}
  These quantities are taken over rectangular domains that share a $t$-axis but, in general, may have different length $s$-axes. In particular, the quantities $\ys$ and $\lambda$ are defined over a common domain $R_S$. The quantities $\yv$ and $\alpha$ are defined over a common domain $R_V$, and the quantities $\yi$, $\gamma$, and $\beta$ are defined over a common domain $R_I$. When these distinctions are not needed, we subsume these s-axes into the common interval $\left[0,\infty\right)$. 
  
  Analogous to \cite{DiLauro2022nonMarkov}, the limiting system can be described as
  \begin{align}
      (\partial_t  + \partial_s)\, y_{S}(t,s) &{} = - \left( \lambda(t,s) + \int_0^\infty \beta(u) y_{I}(t,u) \differential{u}  \right) y_{S}(t,s)\eqcomma \label{eq:ys} \\
      (\partial_t  + \partial_s)\, y_{V}(t,s) &{} = - \left(\alpha(s) + (1-\alpha(s))\int_0^\infty \beta(u) y_{I}(t,u) \differential{u} \right) y_{V}(t,s)\eqcomma \nonumber\\
      (\partial_t  + \partial_s)\, y_{I}(t,s) &{} = - \gamma(s) y_{I}(t,s) \nonumber\\
      \frac{\differential{}}{\differential{t}} y_{R}(t) &{} = \int_0^\infty \left(\alpha(s)y_V(t,s) + \gamma(s)y_{I}(t,s)  \right)\differential{s} \eqcomma \label{eq:yr}
  \end{align}
  with initial and boundary conditions
  \begin{align}
      y_S(t,0) &{} =0 \mbox{, for all } t\geq 0, \label{eq:bdys}\\ y_S(0,s) &{}  = f_S(s) \eqcomma  \nonumber\\
      y_V(t,0) &{} = \int_0^\infty \lambda(t,s)y_{S}(t,s) \differential{s}\label{eq:bdyv}\\  y_V(0,s) &{} =0 \mbox{, for all } s\geq 0\eqcomma  \nonumber\\
      y_I(t, 0) &{} = \int_0^\infty y_{S}(t,s) \int_0^\infty \beta(u) y_{I}(t,u) \differential{u} \differential{s} + \int_0^\infty (1-\alpha(s))y_{V}(t,s) \int_0^\infty \beta(u) y_{I}(t,u) \differential{u} \differential{s} \eqcomma \label{eq:bdyi}\\
      y_I(0,s) &{}  = \rho f_I(s) \eqcomma, \nonumber\\
      y_R(0) &{} = 0\eqstop 
  \end{align}
  We assume there are no vaccinated or removed individuals initially. 
  The nonnegative functions $f_{S}$ and $f_{I}$ describe the initial age distributions of the susceptible and the infected individuals,  and satisfy 
  \begin{align*}
      \int_0^\infty f_{S}(s) \differential{s}  = 1 \eqcomma \quad 
      \int_0^\infty f_{I}(s) \differential{s}  = 1 \eqstop 
  \end{align*}
  The parameter $\rho$ is the initial proportion of infected individuals in the population. It is worthwhile to point out that the system is mass conserved, \ie, 
  \begin{align*}
      \int_0^\infty \left( y_{S}(t, s) + y_{V}(t,s) + y_{I}(t,s) \right) \differential{s}  = 1 + \rho - y_R(t)\eqcomma \quad \text{ for all } t \ge 0 \eqstop 
  \end{align*}
  A consequence of the above conservation law is that the equation \eqref{eq:yr} is in fact redundant. 
      
    \subsection{Continuity constraints on boundary conditions}\label{subsec:cnstbcs}
    When we seek solutions of \eqref{eq:ys}-\eqref{eq:bdyi} in the space of continuous functions, the explicit and implicit boundary data must be assigned continuously at the origin. This criterion is not satisfied freely but imposes additional constraints on the equation coefficients and the initial data. We obtain these constraints by equating the expressions for the explicit and implicit data at the origin and derive
    \begin{align}
        f_S(0) &= 0\eqcomma \label{eq:cnstbcs} \\
        \lambda(0,s) & =  0 \eqcomma  \text{ for all } s \ge 0\eqcomma  \nonumber \\
        f_I(0) &= \int^\infty_0 \beta(u)f_I(u)\differential{u}. \nonumber
    \end{align}
    The last equality generally reduces the combined degrees of freedom for $\beta$ and $f_I$ by one. We note also that the above constraint on $\lambda$ could be relaxed to holding over the support of $f_S$ only.
  
  \subsection{\ac{DSA} perspective of the mean-field \acp{PDE}}\label{subsec:DSA}
  Before describing how we solve the limiting mean-field \ac{PDE} system in the next section, let us briefly discuss how \ac{DSA} interprets the limiting \acp{PDE} as probabilistic quantities.
  
  The \ac{DSA} method \cite{KhudaBukhsh2020InterfaceFocus,DiLauro2022nonMarkov,Wascher2021IDSA,OSU_whitepaper,Harley2022Ebola,KhudaBukhsh2021Prison,Caleb2020Stone,KhudaBukhsh2022Israel} combines classical dynamical systems theory and survival analysis to interpret the mean-field limits of scaled population counts or densities as characterizing probability laws of transfer times between compartments. In the Markovian model, this interpretation boils down to treating the mean-field \acp{ODE} as satisfying a time inhomogeneous Chapman--Kolmogorov equation for the marginal probability law of a Markov chain on the state space $\{\textbf{S}, \textbf{V}, \textbf{I}, \textbf{R} \}$ describing the time-evolving status of a single individual embedded in an infinitely large population. See \cite[Section 3.4]{DiLauro2022nonMarkov} for the standard \ac{SIR} model example. In the non-Markovian case, we  construct a Markov process on the state space  $\{\textbf{S}, \textbf{V}, \textbf{I}, \textbf{R} \}\times [0, \infty)$ to keep track of the time-evolving status as well as the age information of an individual embedded in an infinitely large population. As such, \ac{DSA} interprets the mean-field limiting \acp{PDE} as describing the transition kernel for the Markov process. The transition kernel could be used to simulate individual trajectories. 
  
  Note that the individual-based Markov process (or chain in the Markovian case) is entirely characterized by the mean-field limiting \acp{PDE} (or \acp{ODE} in the Markovian case) describing population-level densities (or counts).  It is precisely in this sense that \ac{DSA} turns an ecological model into an agent-based model! Moreover, this agent-based description gives us the following probability measures
  \begin{align}
      \mu_S(t, A) = \frac{\int_A y_S(t,s) \differential{s} }{ 1+ \rho}\eqcomma \quad  \mu_{V}(t, A) = \frac{\int_A y_V(t,s) \differential{s} }{ 1+ \rho}\eqcomma \quad  \mu_I(t, A) = \frac{\int_A y_I(t,s) \differential{s} }{ 1+ \rho} \eqcomma 
  \end{align}
  for Borel subsets $A$ of $[0, \infty)$. Here, the quantity $\mu_S(t, A)$ describes the probability of a randomly chosen individual with age in the set $A \subset [0, \infty)$ to be in the \textbf{S} compartment at time $t$. The other probability measures are interpreted in a similar fashion. Based on a random sample of infection, vaccination and/or recovery times, which are allowed to be censored, truncated or even aggregated over time or individuals, the above probability measures (and the transition kernel) can be used to write an individual-based product-form likelihood function, called the \ac{DSA}-likelihood. See \cite[Section 3.3]{DiLauro2022nonMarkov} for an explicit example of the \ac{DSA}-likelihood. The \ac{DSA}-likelihood can be used for likelihood-based approaches to parameter inference, such as \ac{MLE}, or Bayesian methods employing \ac{MCMC} techniques. In this paper, we focus on an alternative likelihood-free approach based on the \ac{ABC} method, which we describe in \Cref{sec:num_results}. In the next section, we describe how we solve the mean-field limiting \ac{PDE} system.

  \section{Solving the limiting \ac{PDE} system}\label{sec:PDE_solution}
  
  For the remainder of the paper, we will make a simplifying assumption that the initial data $f_S$ and $f_I$ are compactly supported. This assumption holds in almost all practical cases of interest. For example individuals are of bounded age, and infection lasts for at most a bounded length of time. Moreover, by a suitable enlargement of the rectangular domains of $\ys$, $\yv$, and $\yi$, respectively $R_S$, $R_V$, and $R_I$, we may also assume without loss of generality that the solutions stay compactly supported in their domains for the time horizon simulated. This follows from the method of characteristics used in \Cref{subsec:solmthd} and the forcing terms on the right-hand side of \eqref{eq:ys} which determine exponential solutions. For concreteness, let us write $R_S=\left[0,T\right]\times\left[0,L_S\right]$, $R_V=\left[0,T\right]\times\left[0,L_V\right]$, and $R_I=\left[0,T\right]\times\left[0,L_I\right]$.
  \subsection{Solution method}\label{subsec:solmthd}
  The governing equations, \eqref{eq:ys}-\eqref{eq:bdyi}, constitute a quasi-linear \ac{PDE} system of nonlocal conservation laws. Since the associated directions of their derivative operators are constant and do not intersect, the solutions into the domain interior are naturally suited for construction by the method of characteristics, which recasts the first order \acp{PDE} into equivalent flow equations. For more details on this construction, we refer readers to standard references, such as \cite[Ch 3]{Evans10} and \cite[Ch 7]{DiBenedetto10}.
  
  We note that the implicit and nonlocal boundary conditions, \eqref{eq:bdyv}-\eqref{eq:bdyi}, are not standard. In order to handle the boundary conditions,  we exploit a special feature of this system that remarkably depends on the nonlocal character of the implicit boundary terms. Specifically, we  differentiate the implicit boundary conditions in $t$ and recover a governing flow for them, which only depends on the solution and not on the solution's derivatives. For example, consider differentiating \eqref{eq:bdyv}:
  \begin{align*}
  \partial_t\yv(t,0) &= \int^{L_S}_0\left(\partial_t\lambda\ys + \lambda\partial_t\ys\right)ds = \int^{L_S}_0\left( \partial_t\lambda\ys+\lambda\left(-\partial_s\ys-\lambda\ys -\ys\int^{L_I}_0\beta(u)\yi(t,u)du \right)\right)ds\\
  &= \left.-\left[\lambda\ys\right](t,s)\big|^{L_S}_{s=0}\right. + \int^{L_S}_0\left[\left(\partial_t\lambda+\partial_s\lambda\right)\ys\right](t,s) ds - \int^{L_S}_0\left[\lambda^2\ys\right](t,s)ds \\
  &\;\;\;\;- \left(\int^{L_S}_0\left[\lambda\ys\right](t,s)ds \right)\left(\int^{L_I}_0\left[\beta\yi\right](t,s)ds\right).
  \end{align*}
  In the second equality, we have substituted for $\partial_t\ys$ using the $\ys$ equation of \eqref{eq:ys}, and in the third equality, we performed an integration by parts. Continuing we may use the compactly supported initial data to conclude that the boundary term at $s=L_S$ is vanishing and substitute \eqref{eq:bdys} into the boundary term at $s=0$. For the particular case of $\yv$, we see both these terms are vanishing. A similar argument works for $\partial_t\yi(t,0)$. Below we present the resulting evolution equations for the solution's boundary values. Of course, $\ys(t,0)\equiv 0$ by \eqref{eq:bdys}, so we need only consider the remaining two:
  \begin{align}
  \partial_t\yv(t,0) &= \int^{L_S}_0\ys(\dcl)\lambda  ds - \int^{L_S}_0 \lambda^2\ys ds - \left(\int^{L_S}_0 \lambda\ys ds\right)\left(\int^{L_I}_0 \beta\yi ds\right)\eqcomma\label{eq:bdflow_yv}\\
  \partial_t\yi(t,0) &= 
  \left(\int^{L_I}_0\beta\yi ds\right)\left[-\int^{L_S}_0\lambda\ys ds - \left(\int^{L_S}_0\ys ds\right)\left(\int^{L_I}_0\beta\yi ds\right)-\int^{L_V}_0\yv(\dt +\ds)\alpha ds\right.\nonumber\\ 
		&\left. \;\;\;\;\;\; +\left[\big(1-\alpha(t,0)\big)\yv(t,0)\right] -\int^{L_V}_0 \left(1-\alpha\right)\alpha\yv ds- \left(\int^{L_V}_0\left(1-\alpha\right)^2\yv ds\right)\left(\int^{L_I}_0\beta\yi ds\right)\right]\nonumber\\
		&\;\;+\left(\int^{L_S}_0\ys ds+\int^{L_V}_0(1-\alpha)\yv ds\right)\left(\int^{L_I}_0\yi(\dt+\ds)\beta ds + \beta(0)\yi(t,0) - \int^{L_I}_0\beta\gamma\yi ds  \right).\nonumber
  \end{align}
  Note that in \eqref{eq:bdflow_yv}, it is understood that all integrands are evaluated at $(t,s)$ and integrated in $s$. In our particular application, we also have $\left(\dcl\right)\alpha=\partial_s\alpha$ and similarly for the hazard function $\beta$. We have chosen to present \eqref{eq:bdflow_yv} in generality to highlight the underlying structure of \eqref{eq:ys}-\eqref{eq:bdyi}. Note also, the validity of the equations \eqref{eq:bdflow_yv} depends on the compact support of the initial data. Otherwise, these flow equations would need to contain additional boundary terms at the right-end points of the s-axes. 
  
  Our solution method of \eqref{eq:ys}-\eqref{eq:bdyi} now consists of coupling \eqref{eq:ys} in the interior domain together with the alternative flow formulation \eqref{eq:bdflow_yv} of the implicit boundary conditions at the $\left[s=0\right]$ boundary. The explicit boundary data assigned at $\left[t=0\right]$ in \eqref{eq:bdys}-\eqref{eq:bdyi} becomes this system's initial data. For the benefit of readers mostly interested in applications, we defer a rigorous analysis of this alternative formulation and its equivalence with the original one to Appendix~\ref{app:subsecflow}. Also, in this investigation, we assume equation coefficients in classical function spaces and classical notions of solvability. Future work may investigate  deeper questions of equivalence between the original and flow formulations when equation coefficients and the solution belong to suitably identified Sobolev spaces. For more information on the distinctions between classical solutions and weak solutions in Sobolev spaces, we refer again to \cite{Evans10,DiBenedetto10}. We present a numerical demonstration of this method approximating the implicit boundary conditions, \eqref{eq:bdyv}-\eqref{eq:bdyi}, in \Cref{appsubsec:bdconv}.
  
  \subsection{Numerical implementation}
  For numerically approximating \eqref{eq:ys} and \eqref{eq:bdflow_yv}, we take a semi-discrete approach where we discretize the space axis, $s$, into a number of uniform intervals whose intersection points are called nodes. The number of nodes determines the mesh resolution. We then approximate each component of the true solution -- $\ys$, $\yv$, and $\yi$ -- by its own nodal basis expansion
  \[
  y(t,s) \approx \sum^n_{i=1} y_i(t)\phi_i(s).
  \]
  The $\left\{\phi_i(s)\right\}_i$ functions are piecewise linear, $C^0$ basis splines determined by a given mesh resolution. Each basis spline is associated with an $s$-axis node, $s_i$, and is defined by taking the value 1 at that node and 0 at all others. At each $t$-level, we thus approximate the true solution with a linear, interpolant spline whose value at $(t,s_i)$ is given by $y_i(t)$. Integrals of the true solution and its derived quantities are approximated by taking integrals of their corresponding linear, interpolant splines by trapezoidal rule.
  
  The $\left\{y_i(t)\right\}_i$ evolve according to the flows of \eqref{eq:ys} and  \eqref{eq:bdflow_yv}. We initialize at the $t=0$ level by projecting the initial data onto the interpolant basis. This is accomplished by sampling those functions at the nodes. We then use an explicit Euler scheme to extend the solution up to the next time step, $t+\Delta t$ along the boundary axis where the implicit data is prescribed according to the \acp{ODE} of \eqref{eq:bdflow_yv}. This determines the first value $y_1(t+\Delta t)$ at the node $s=0$.  For the remaining nodal values, each of their nodes originate along a characteristic from either the previous $t$-level or within that part of the boundary axis that was previously approximated in computing $y_1$.  We now evolve each of those originating, solution values by the method of characteristics along the \acp{ODE} that correspond to \eqref{eq:ys} by a second explicit Euler scheme to obtain the remaining $\left\{y_i(t+\Delta t)\right\}^n_{i=2}$. By iterating this process, we construct a numerical approximation over the desired time interval for simulation. 
  
  In practice, we use an adaptive Euler step to improve solution accuracy. We specify tolerances atol and rtol, which are respectively the absolute and the relative tolerances set by the user. The solver approximates the solution using a single Euler step of size $\Delta t$ and two consecutive smaller Euler steps of size $\Delta t/2$. The solver then checks if at each node the difference between the two Euler approximations satisfies either the atol or rtol thresholds. It does this similarly for the values of the nonlocal integrals. It accepts the step if all quantities meet this criterion, or else it halves the step size and repeats this process. Note that in evaluating the absolute error, we take the magnitude difference in the Euler solutions and then scale by the length of their $s$-axes. This is done because, in general, $\ys$, $\yv$, and $\yi$ are defined on $s$-axes of different lengths, while probability densities vary inversely with the length of their domain. By this adjustment, we normalize the errors in $\ys$, $\yv$, and $\yi$ to a common scale. 
  
  Below we also provide a pseudocode of the algorithm described in this section.
  \begin{pseudocode} Numerical solver for \eqref{eq:ys} and \eqref{eq:bdflow_yv}
  \begin{enumerate}
      \item Initialize the solution at $t=0$ by storing values of the initial data at the nodes,
      \item Propagate from $y_1(t)$ to $y_1(t+\Delta t)$ by taking an explicit Euler step of size $\Delta t$ along \eqref{eq:bdflow_yv},
      \item Propagate from $\left\{y_i(t)\right\}^n_{i=1}$ and $y_1(t+\Delta t)$ to $\left\{y_i(t+\Delta t)\right\}^n_{i=2}$ by taking Euler steps of size at most $\sqrt{2}\Delta t$ along the characteristics of \eqref{eq:ys},
      \item Repeat Steps 2-3 for two consecutive Euler steps with $\Delta t = \Delta t/2$,
      \item If the y-nodal values and integral quantities computed by the single and two-half Euler steps satisfy either of the absolute and relative error thresholds, atol and rtol, then accept the Step 4 solution at $t+\Delta t$ by storing its nodals and continue with $\Delta t = 2\Delta t$. Otherwise, repeat steps 2-4 with $\Delta t=\Delta t/2$.
  \end{enumerate}
  \end{pseudocode}
  
  \subsubsection*{Code availability}
  We provide an implementation of the algorithms described in this paper in the Julia programming language along with jupyter notebooks and html at \cite{github}. Additional information on Julia may be found at \cite{Julia2017}.

  \section{Model parameters}\label{sec:prm}
  
  For the vaccine efficacy, $\alpha$, we begin with a function that linearly increases from 0 to a maximum of $\aeff$ over a period of $\alpha_L$ days. After $\alpha_L$, it was taken to be constant. For the contact interval and infectious period hazard functions, $\beta$ and $\gamma$, we choose Weibull distributions as the Weibull distribution has been shown to be a flexible choice in modelling epidemics \cite{DiLauro2022nonMarkov}. Therefore, we have
    \begin{align*}
      \alpha(s) &{} \defeq \frac{\min (s, \alpha_L)}{\alpha_L} \times \aeff \eqcomma \quad  s\ge 0 \eqcomma \\
      \beta(s) &{} \defeq \frac{\beta_\alpha}{\beta_\theta} \left(\frac{s}{\beta_\theta} \right)^{\beta_\alpha-1} \eqcomma \quad  s \ge 0 \eqcomma \\
      \gamma(s) &{} \defeq \frac{\gamma_\alpha}{\gamma_\theta} \left(\frac{s}{\gamma_\theta} \right)^{\gamma_\alpha-1} \eqcomma \quad s \ge 0 \eqstop 
    \end{align*}
  
  \noindent Further, we smoothed $\alpha$ by a moving integral average to ensure its derivative was everywhere classical. We took the density for the initial infected population, $f_I$, to be approximately uniform over a period of fourteen days subject to the further requirement that it should be compactly supported and continuous. For this purpose, $f_I$ was constant up to day 13.25 and then linearly decayed to 0 by day 13.75. As with $\alpha$, we  smoothed $f_I$ by an integral moving average. 
  
  We inferred the parameters in \eqref{eq:ys}-\eqref{eq:bdyi} using COVID-19 epidemic data for the US state of Ohio, \cite{ODH21}, during the time period Nov 15, 2020- Jan 15, 2021. We directly estimated the hazard rate for vaccination for this same time period from the \ac{ODH} and \ac{CDC} data, \cite{ODH21,CDC21}, using b-splines. We also directly estimated the density for the initial susceptible population, $f_S$, consistent with US census information on age distributions in Ohio, \cite{USCEN21}. 
  \subsection{Empirical estimates}\label{subsec:prm_data}
  \subsubsection{The distribution for the initial susceptible population, $f_S$}
\begin{figure}[h!]
      \centering
      \includegraphics[scale=1]{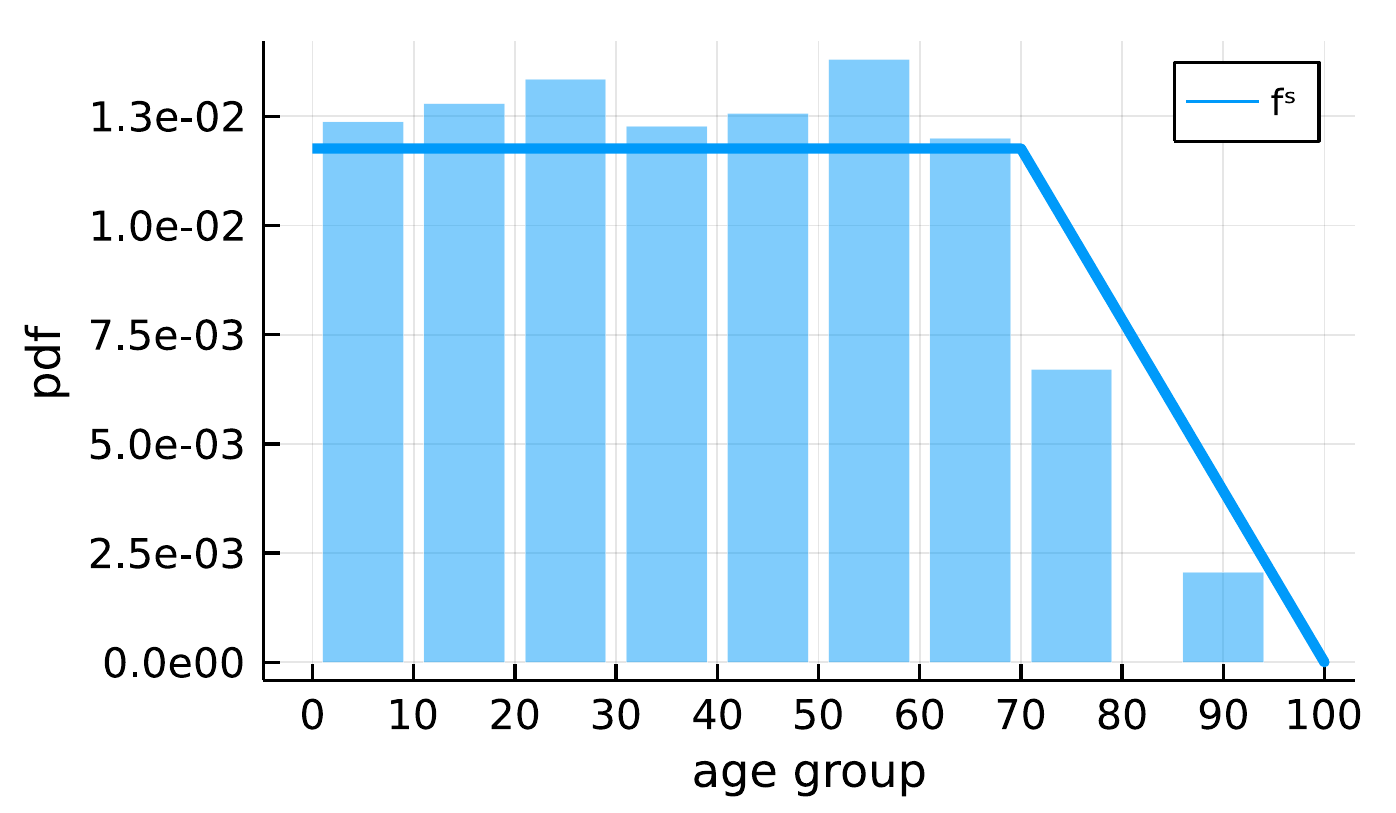}
      \caption{\label{fig:fs}%
      An empirical density for the age distributions in Ohio partitioned into ten year brackets. From this we extracted a probability density curve, $f_S$. Later we additionally make this density decay to 0 at the origin consistent with \eqref{eq:cnstbcs}. 
      } 
  \end{figure}

  \Cref{fig:fs} shows an empirical density for age distributions in the state of Ohio partitioned into ten year age groups. This data suggested a simple piecewise linear characterization of $f_S$, which is also shown. Additionally, we normalize $f_S$ so that it integrates to one.
  \subsubsection{The hazard rate for time to vaccination $\lambda$}
\begin{figure}
      \centering
      \includegraphics[scale=0.9]{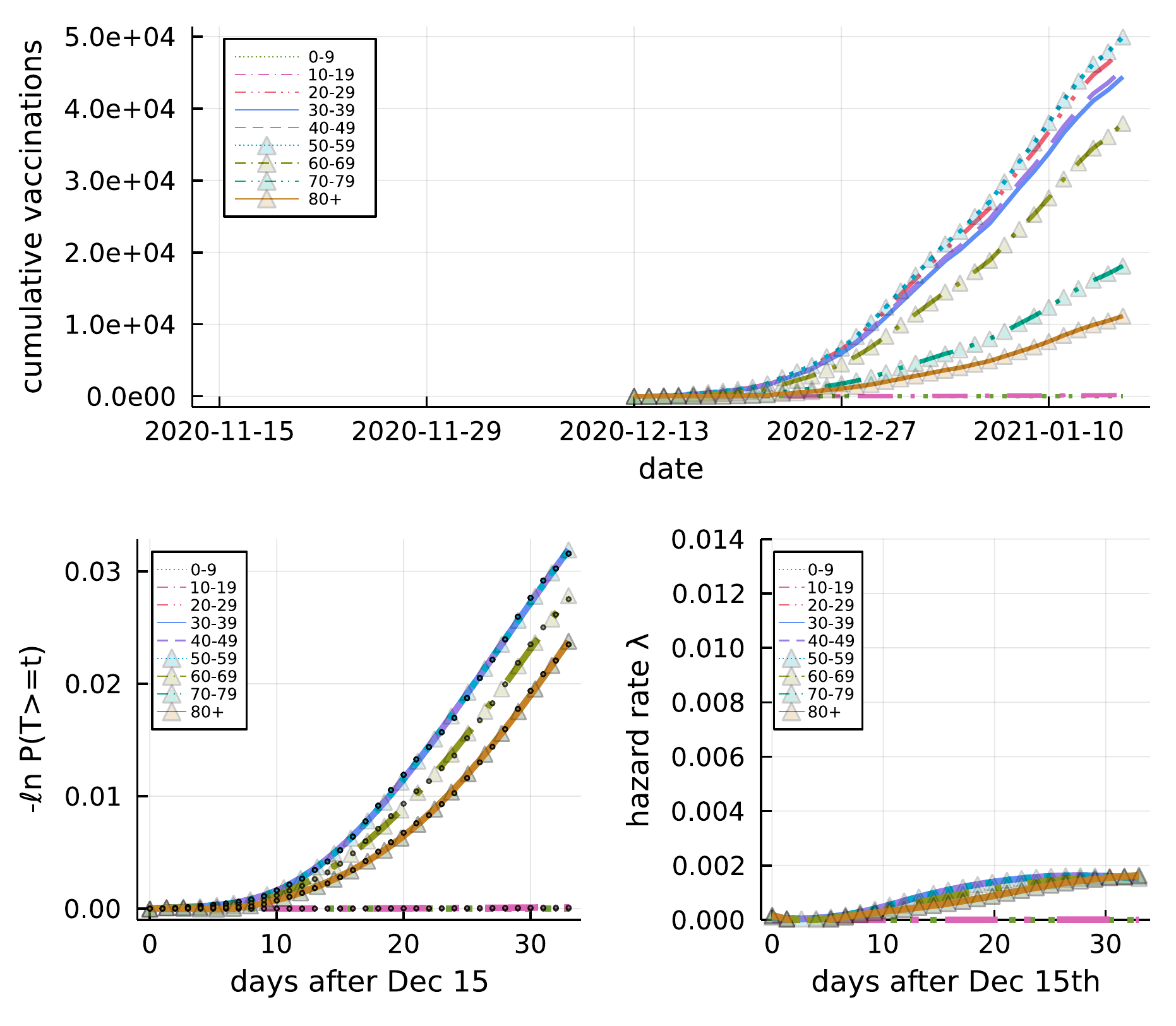}
      \caption{\label{fig:mst_vax}%
      \textit{(Top)} Cumulative vaccine doses administered in Ohio. \textit{(Bottom left)} From the cumulative vaccine doses and the Ohio population's age distribution, we log transformed the empirical vaccine survival function. The empirical data is shown as dots. These points were then fit by least-squares cubic splines joined together with $C^2$ smoothness which are also shown. The slopes of these curves represent the vaccine hazard rate. \textit{(Bottom right)} The nonparametric hazard rates, computed as the derivatives of the spline curves in the adjacent panel, are shown.}
      
  \end{figure}

  \Cref{fig:mst_vax} shows the empirical, cumulative vaccination doses given in Ohio from Nov 15th, 2020 - Jan 15th, 2021 (top panel) broken down by age group. These age-breakdowns were estimated from the vaccination counts provided by the \ac{ODH} and \ac{CDC}, \cite{ODH21, CDC21}. We observe that vaccine rollout did not begin until Dec 15th. It also shows the derived empirical estimations of the hazard rate for vaccination, $\lambda$, from this data (bottom two panels).  In order to do this, we measured an empirical survival function for vaccination using the cumulative vaccine doses curves and the age demography in Ohio and then applied to these values the negative $\ln$-transformation. These data points were then fit by least squares with $C^2$ cubic splines, and their pointwise derivatives corresponded to the desired vaccination hazard rate. For more information on computing with splines, we refer to \cite{Schumaker15}. These several curves, parameterized in $t$, were combined into a single $\lambda(t,s)$ function by using smoothed linear transitions in the age-variable $s$ across age-groups. 
  \subsection{Estimation by \acl{ABC}}\label{subsec:prm_abc}
  The remaining model parameters were not explicitly given by data, and we used instead an \ac{ABC} scheme to estimate their values. The goal of Bayesian inference is to obtain samples from the posterior distribution of the parameters, which is proportional to a prior distribution on the parameters times the likelihood of the data given the parameters. However, in many cases, it is difficult or expensive to compute or even sample from the posterior. In particular, many methods for sampling from the posterior require computing the likelihood, which may itself be intractable.
  
  \ac{ABC} provides a method to approximate posterior samples without computing the likelihood. The simplest way to perform \ac{ABC} is the rejection sampling method, which goes as follows: First, one proposes a vector of parameters from the joint prior. Next, one simulates data (or some summary statistic thereof) from the model using this vector of parameters. Finally, one compares this simulated data to the observed data using some distance metric and either accepts the proposed vector of parameters as a sample from an approximation of the posterior or rejects it depending on how close it is to the observed data. What is sufficiently close for acceptance may be determined by an absolute threshold or by retaining some proportion of the best samples. The intuition is that a vector of parameters with higher likelihood will more often produce simulated data that is close to the empirical data, and so \ac{ABC} approximates the usual acceptance procedure based on the likelihood ratio that is often used in MCMC. For a more thorough discussion of \ac{ABC}, we direct the reader to \cite{kypraios2017tutorial,ABC}.
  
  We proposed the vector of parameters $(\beta_{\alpha}, \gamma_{\alpha}, \gamma_{\theta}, \rho, \alpha_{\textrm{eff}})$ by drawing each quantity independently from a uniform prior with the bounds given in Table \ref{tab:prms}. We then accepted it if the associated infectious period was less than three weeks with 99.9\% probability and if the mean contact interval was less than the mean infectious period. The parameter $\alpha_L$ was fixed to be two weeks. The parameter $\beta_{\theta}$ was not proposed, and hence, does not have a prior, since its value was determined from the other parameters. This followed from the last constraint in \eqref{eq:cnstbcs}. In fact, for the particular pairing of a Weibull distribution for the contact interval and a uniform distribution for the initial infected, it follows from a little algebra that \eqref{eq:cnstbcs} implies $\beta_\theta$ is the length of the support of $f_I$ and independent of the other parameters. 
  
  We then used the \ac{PDE} model to generate a predicted incidence trajectory for the Nov 15, 2020 - Jan 15, 2021 time period and compared the predicted trajectory to the empirical trajectory using the root mean square error (\ac{RMSE}). Note that the prevalence at time $t$ is given by $n \int_0^\infty y_{I}(t,s) \differential{s}$, where $n$ is the total population size.  We generated 5000 \ac{ABC} sample trajectories in this fashion and retained the $10\%$ of sample trajectories with the lowest \ac{RMSE}. The results are shown in  \Cref{fig:abcvar}.

  \section{Numerical results}\label{sec:num_results}
  
  Here we demonstrate how the \ac{PDE}-\ac{DSA} model can be used in conjunction with with the \ac{ABC} method to infer model parameters from epidemic data. As discussed in \Cref{subsec:prm_abc}, the \ac{ABC} method does not require an explicit likelihood but instead uses a computable error between synthetic and true data values to assess the quality of proposed parameter values. Aggregate population-level counts of infection are prototypical data that would be used with \ac{DSA}, and so we apply this approach to daily reported incidence supplied by the ODH during the period of  Nov 15, 2020, - Jan 15, 2021. This period encompassed the epidemic wave promoted by the rise of the COVID-19 $\alpha$-variant \cite{CDC22} as well as the start of vaccine roll-out in Ohio. 
  
\begin{figure}[h!]
      \centering
      \includegraphics{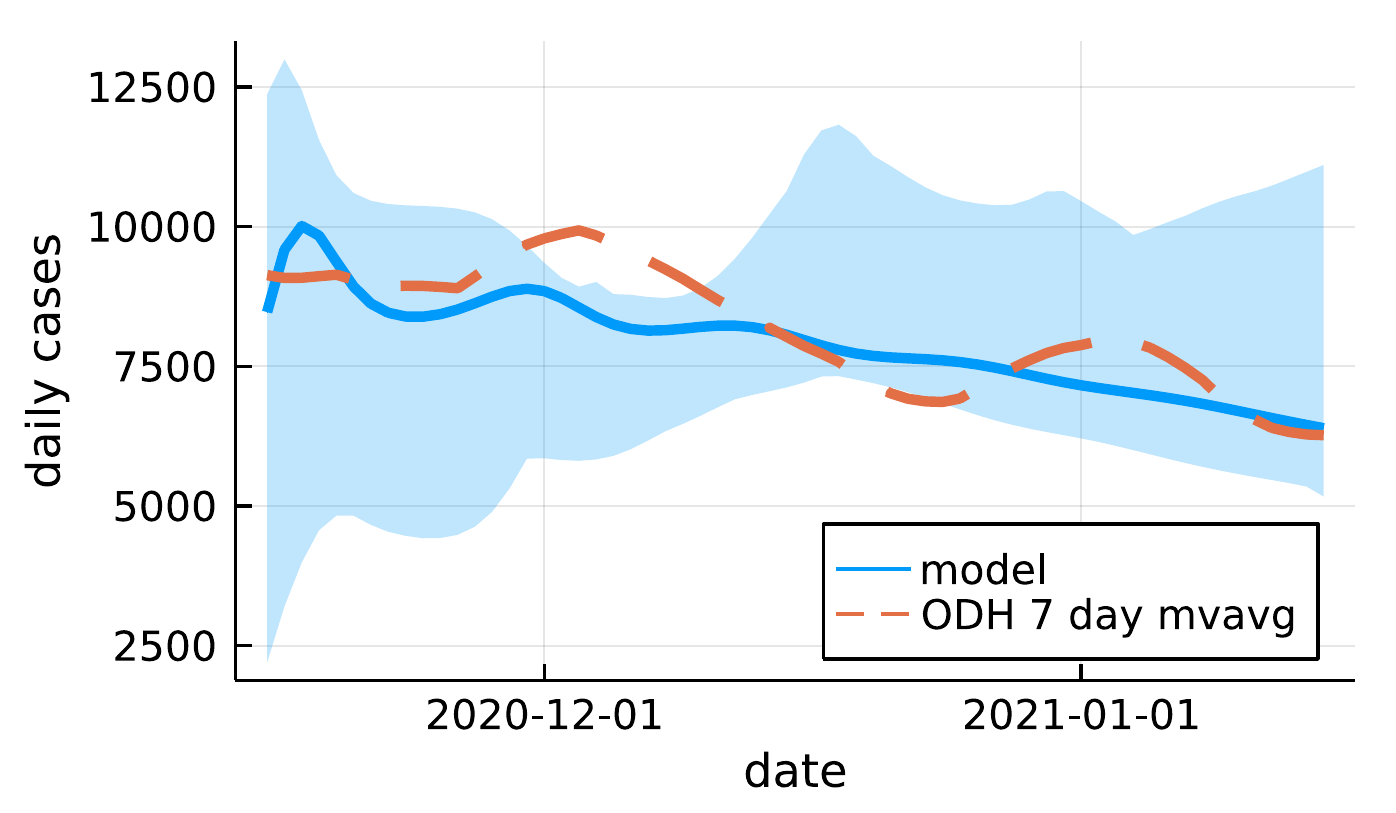}
      \caption{The \ac{PDE} model fit to the \ac{ODH} daily incidence data as calibrated by \ac{ABC}. A seven day moving average of the \ac{ODH} data is shown as a dashed orange line. The solid blue line corresponds to the model's best fit observed across 5000 \ac{ABC} samples, while the blue band represents the best 10\% of all \ac{ABC} sampled trajectories as measured by the \ac{RMSE} between the model prediction and empirical incidence.}
      \label{fig:abcvar}
\end{figure}

\begin{figure}[t!]
      \centering
      \includegraphics[scale=0.9]{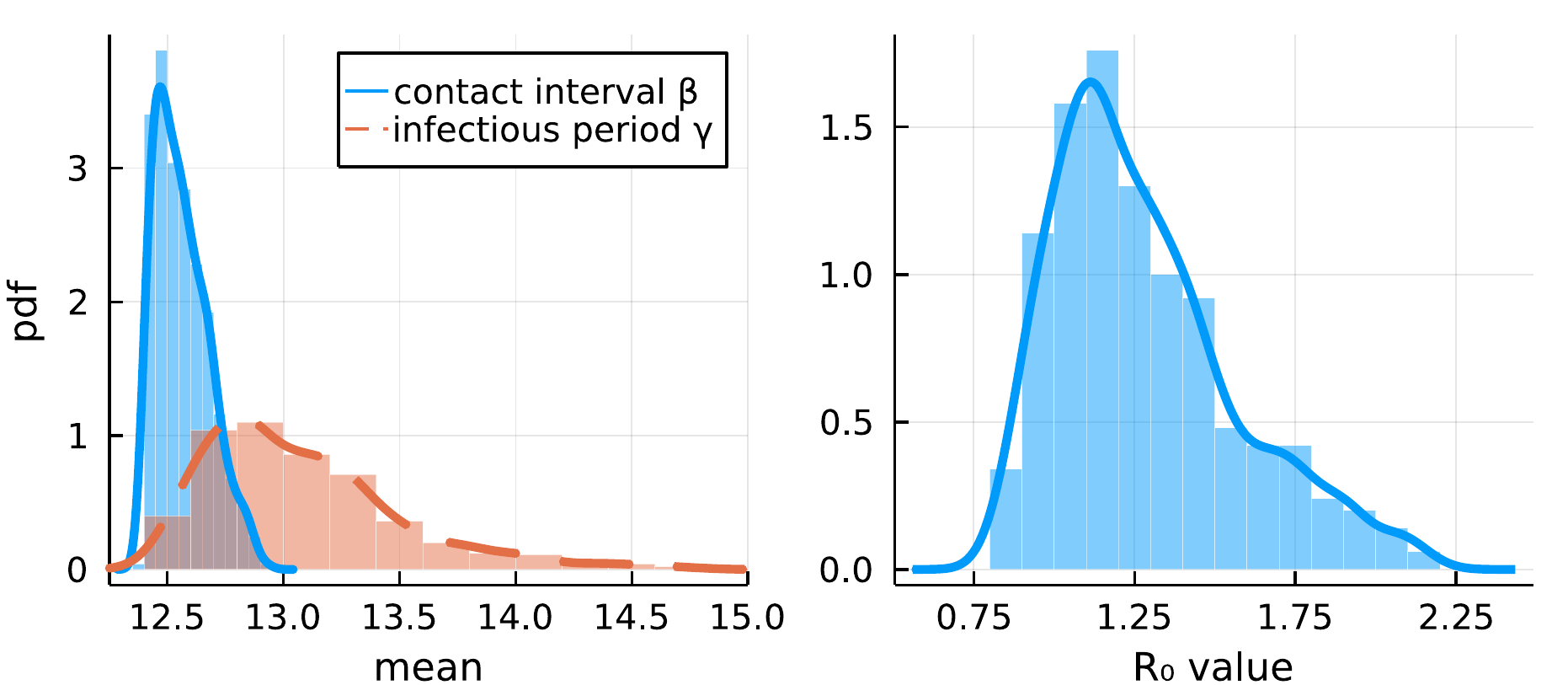}
      \caption{\textit{(Left)} Posterior distribution of the mean contact interval compared with the posterior distribution for the mean infectious period. \textit{(Right)} Posterior distribution for $R_0$. }
      \label{fig:mst_meanR0}
  \end{figure}

  The primary parameters of interest were for the contact interval characterized by $\beta$ and the infectious period characterized by $\gamma$, along with estimates of the initial amount of infection $\rho$. The vaccine efficacy parameter $\aeff$, while also relevant, would not be identifiable due to still low rate of vaccine administration  during  the time window  of interest. \Cref{fig:abcvar} shows the model predictions  together with the \ac{ODH} reported daily incidence.  The best fit obtained across all 5000 \ac{ABC} samples captures the nontrivial empirical trends.  \Cref{tab:prms} lists the model parameters with their best fit values and the \ac{ABC} posterior credible intervals. The posterior distributions correspond to the parameter values that produced trajectories within the shaded bands of \Cref{fig:abcvar}. In \Cref{fig:mst_meanR0}, we see that the mean time before transmission was approximately estimated to be between $\left(12.25,13.0\right)$ days while the mean time to recovery was approximately estimated to be between $\left(12.25,15.0\right)$ days. These estimates are in line with estimates from  similar studies reported  in the literature  \cite{DiLauro2022nonMarkov}.  
  
\noindent In the second panel of \Cref{fig:mst_meanR0}, we give the posterior distribution for the basic reproduction number $R_0$. This quantity is computed using the formula $R_0=\int^\infty_0 S_{\gamma}(s)\beta(s)\differential{s}$ as in \cite{KhudaBukhsh2020InterfaceFocus}, where $S_\gamma(s)$ is the survival function of the probability distribution characterized by the hazard rate $\gamma$. Note that this formula ignores vaccination, and therefore, is an upper bound on the true basic reproduction number. The model estimated a mean of $R_0=1.28$ with a 95\% credible interval of $\left(0.89,1.94\right)$, which is consistent with other studies of basic reproduction numbers associated with the COVID-19.
  \begin{table}[h!]
      \centering
      \begin{tabular}{|c|c|c|c|c|c|}
           \hline
           Parameter & Unit & Description & 
           \begin{tabular}{@{}c@{}}
                Value/  \\
                Prior
           \end{tabular} & \begin{tabular}{@{}c@{}}Best \\Fit\end{tabular} & 
           \begin{tabular}{@{}c@{}}
                \ac{ABC} Posterior \\
                95\% Credible Interval
           \end{tabular}\\
           \hline
           \multicolumn{6}{|c|}{\textbf{Contact interval}}\\
           \hline
           $\beta_\alpha$ & -- & Weibull shape parameter & $\left(5,20\right)^*$ & 5.24 & $\left(5.06,9.55\right)$\\
           \hline
           $\beta_\theta$ & day & Weibull scale parameter & -- & 13.5038 & $\left( 13.5036,13.5076\right)$\\
           \hline
           \multicolumn{6}{|c|}{\textbf{Infectious period}}
           \\
           \hline 
           $\gamma_\alpha$ & -- & Weibull shape parameter & $\left(2,8\right)^\ast$ & 6.05 & $\left(4.98,7.94\right)$\\
           \hline
           $\gamma_\theta$ & day & Weibull scale parameter & $\left(1,18\right)^*$ & 13.6 & $\left(13.43,15.03\right)$\\
           \hline
           \multicolumn{6}{|c|}{\textbf{Starting infection}}
           \\
           \hline
           $\rho$ & \% & 
           \begin{tabular}{@{}c@{}}
                mass of the initial \\
                infected population \\
                relative the size of the \\
                susceptible population
           \end{tabular}
           & $\left(0.1,5\right)$ & 4.48 & $\left(0.79,4.95\right)$
           \\
           \hline
           \multicolumn{6}{|c|}{\textbf{Vaccine}}
           \\
           \hline
           $\alpha_L$ & day & \begin{tabular}{@{}c@{}}
                time for vaccine to achieve  \\
                full efficacy  
           \end{tabular} & 14 & -- & --\\
           \hline
           $\alpha_{\mathrm{eff}}$ & \% & vaccine blocking efficacy & $\left(70,100\right)$ & 97.0 & $\left(70.7,99.2\right)$\\
           \hline
      \end{tabular}
      \caption{Model parameters used to fit the \ac{ODH} daily case data from Nov 15, 2020 - Jan 15, 2021. We additionally report their assigned \ac{ABC} priors, best fit values, and \ac{ABC} posterior credible intervals from the numerical simulations presented in \Cref{sec:num_results}. The \ac{ABC} posteriors were obtained by conditioning 5000 \ac{ABC} samples to the best 10\% of observed \ac{RMSE} between the \ac{ODH} daily incidence and model prediction. The priors for $\beta$ and $\gamma$ are listed with * since, in addition to these uniform ranges, we further conditioned on parameters where the infectious period lasted no more than 21 days with 99.9\% probability and where the mean contact interval was less than the mean infectious period. Moreover, a prior for $\beta_\theta$ is not given since \eqref{eq:cnstbcs} determines $\beta_\theta=\beta_\theta(\beta_\alpha,f_I)$. The narrow posterior of $\beta_\theta$ was expected as reasoned in \Cref{subsec:prm_abc}.}
      \label{tab:prms}
  \end{table}

  \section{Discussion}\label{sec:discussion}
  



The COVID-19 pandemic has spurred the development of a cottage industry of ever-more elaborate mathematical models  of epidemic dynamics that have been applied to   empirical data across  the world with varying degree of success \cite{Koelle22,Driscoll21, Holmdahl20,Jewel20,Barda20}.  Since most of the current and historic  COVID-19 data have been   available in aggregate, the  models tend to focus on  aggregate behavior which may sometimes lead to erroneous insights \cite{klaus2022assortative}.  The \ac{DSA} method  discussed here offers a different modeling approach  and in particular  accounts more fully that a typical compartmental model  for  the heterogeneity of individual  behaviors. 

Indeed, even since its introduction in \cite{KhudaBukhsh2020InterfaceFocus} on the eve of an outbreak of the global  COVID-19 pandemic, the \ac{DSA} approach has been shown to be  a viable  way of analyzing epidemic data  in order to predict epidemic progression, evaluate the long term  effects  of public health policies (testing, vaccination, lock downs, etc.) and  individual-level  decisions (masking, social distancing, vaccine hesitancy) \cite{KhudaBukhsh2022Israel, Harley2022Ebola,KhudaBukhsh2021Prison,Wascher2021IDSA,OSU_whitepaper}.   
As it was argued in  \cite{KhudaBukhsh2020InterfaceFocus}, the \ac{DSA} method has several advantages over  more traditional  approaches  to  analyzing data in SIR-type epidemic curves. Perhaps the most important one is  that the   method does not require the knowledge of the full curve trajectory or the size of the total susceptible population. Indeed, such quantities are rarely known in practice and often require ad-hoc adjustments leading to severely biased analysis \cite{kenah2011biostat}. 

In this work we  discuss a relatively simple  yet powerful non-Markovian extension of the original \ac{DSA} method formulation  introduced in
\cite{KhudaBukhsh2020InterfaceFocus}.  This  extension allows one to take into account the  additional heterogeneity of the transmission patterns due  to both the changes  in infectiousness  over the individuals' infectious periods  and the changes in  immunity over the individuals' periods of vaccine-derived protection.   
However, the practical price to pay for these modeling improvements is that a more  elaborate numerical scheme is required to evaluate the \ac{DSA} model along with its increased computational cost to fit empirical data.   Here   we have chosen  to apply a likelihood-free \ac{ABC} approach,  which allows us to avoid the large numerical overhead usually associated with  \ac{DSA} likelihood-based methods \cite{DiLauro2022nonMarkov}. This is accomplished by pregenerating parameter samples and then running simulations in parallel using  modern multi-core capabilities. This capacity of ABC for parallel processing  helps also mitigate the computational burden of a  non-Markovian \ac{DSA}, namely evaluating its nonlocal and implicit  boundary conditions and associated flow formulation of  \eqref{eq:bdflow_yv}. The analytic properties of \eqref{eq:ys} and \eqref{eq:bdflow_yv} are of intrinsic interest in their own right, as  a deeper understanding of solution regularity could help identify numerical schemes of higher and optimal convergence order. We hope to further pursue these lines of investigation in the future. As seen from the numerical examples in \Cref{sec:num_results}, using \ac{ABC} we were able to fit the highly irregular model of Ohio COVID-19 epidemic  with proper accounting  for various sources of uncertainty in all of the relevant  model parameters.

  \appendix
  \section{Acronyms}\label{sec:acronyms}
  
\begin{acronym}[OWL-QN]
    \acro{ABC}{Approximate Bayesian Computation}
	\acro{ABM}{Agent-based Model}
	\acro{BA}{Barab\'asi-Albert}
	\acro{BD}{Birth-Death}
	\acro{CDC}{Centers for Disease Control and Prevention}
	\acro{CDF}{Cumulative Distribution Function}
	\acro{CLT}{Central Limit Theorem}
	\acro{CM}{Configuration Model}
	\acro{CME}{Chemical Master Equation}
	\acro{CRM}{Conditional Random Measure}
	\acro{CRN}{Chemical Reaction Network}
	\acro{CTBN}{Continuous Time Bayesian Network}
	\acro{CTMC}{Continuous Time Markov Chain}
	\acro{DSA}{Dynamical Survival Analysis}
	\acro{DTMC}{Discrete Time Markov Chain}
	\acro{DRC}{Democratic Republic of Congo}
	\acro{ER}{Erd\"{o}s-R\'{e}nyi}
	\acro{ESI}{Enzyme-Substrate-Inhibitor}
	\acro{FCLT}{Functional Central Limit Theorem}
	\acro{FLLN}{Functional Law of Large Numbers}
	\acrodefplural{FLLN}[FLLNs]{Functional Laws of Large Numbers}
	\acro{HJB}{Hamilton–Jacobi–Bellman}
	\acro{iid}{independent and identically distributed}
	\acro{IPS}{Interacting Particle System}
	\acro{KL}{Kullback-Leibler}
	\acro{LDP}{Large Deviations Principle}
	\acro{LLN}{Law of Large Numbers}
	\acrodefplural{LLN}[LLNs]{Laws of Large Numbers}
	\acro{LNA}{Linear Noise Approximation}
	\acro{MAPK}{Mitogen-activated Protein Kinase}
	\acro{MCMC}{Markov Chain Monte Carlo}
	\acro{MGF}{Moment Generating Function}
	\acro{MLE}{Maximum Likelihood Estimate}
	\acro{MM}{Michaelis-Menten}
	\acro{MPI}{Message Passing Interface}
	\acro{MSE}{Mean Squared Error}
	\acro{ODE}{Ordinary Differential Equation}
	\acro{ODH}{Ohio Department of Health}
	\acro{PDE}{Partial Differential Equation}
	\acro{PDF}{Probability Density Function}
	\acro{PGF}{Probability Generating Function}
	\acro{PGM}{Probabilistic Graphical Model}
	\acro{PMF}{Probability Mass Function}
	\acro{psd}{positive semi-definite}
	\acro{PT}{Poisson-type}
	\acro{QSSA}{Quasi-Steady State Approximation}
	\acro{rQSSA}{reversible QSSA}
	\acro{RMSE}{Root Mean Square Error}
	\acro{SD}{Standard Deviation}
	\acro{SEIR}{Susceptible-Exposed-Infected-Recovered}
	\acro{SI}{Susceptible-Infected}
	\acro{SIR}{Susceptible-Infected-Recovered}
	\acro{SIS}{Susceptible-Infected-Susceptible}
	\acro{sQSSA}{standard QSSA}
	\acro{ssLNA}{Slow-scale Linear Noise Approximation}
	\acro{tQSSA}{total QSSA}
	\acro{WS}{Watts-Strogatz}
	\acro{whp}{with high probability}
\end{acronym}

  \section*{Acknowledgments}
  WRK acknowledges financial support from the London Mathematical Society (LMS) under a Scheme 4 grant (Ref No: 42118). GAR acknowledges support from the National Science Foundation under grants  DMS-2027001  and DMS-1853587.

  \section*{Conflict of interest}
  The authors declare no conflict of interest. 

  \bibliographystyle{AIMS}

\begin{thebibliography}{10}

\bibitem{Barda20}
\newblock N.~Barda, D.~Riesel, A.~Akriv, J.~Levy, U.~Finkel, G.~Yona,
  D.~Greenfeld, S.~Sheiba, J.~Somer, E.~Bachmat, G.~N. Rothblum, U.~Shalit,
  D.~Netzer, R.~Balicer and N.~Dagan,
\newblock {{D}eveloping a {C}{O}{V}{I}{D}-19 mortality risk prediction model
  when individual-level data are not available},
\newblock \emph{Nat Commun}, \textbf{11} (2020), 4439.

\bibitem{Caleb2020Stone}
\newblock C.~D. Bastian and G.~A. Rempala,
\newblock Throwing stones and collecting bones: Looking for poisson-like random
  measures,
\newblock \emph{Mathematical Methods in the Applied Sciences}, \textbf{43}
  (2020), 4658--4668.

\bibitem{Julia2017}
\newblock J.~Bezanson, A.~Edelman, S.~Karpinski and V.~B. Shah,
\newblock Julia: A fresh approach to numerical computing,
\newblock \emph{SIAM {R}eview}, \textbf{59} (2017), 65--98.

\bibitem{USCEN21}
\newblock U.~Census,
\newblock {County Population Totals 2010-2019},
\newblock
  \urlprefix\url{{https://www.census.gov/data/datasets/time-series/demo/popest/2010s-counties-total.html}},
\newblock Accessed: September 13, 2021.

\bibitem{choi2019modeling}
\newblock B.~Choi, S.~Busch, D.~Kazadi, B.~Ilunga, E.~Okitolonda, Y.~Dai,
  R.~Lumpkin, O.~Saucedo, W.~R. KhudaBukhsh, J.~Tien, E.~Kenah and G.~Rempala,
\newblock {Modeling outbreak data: Analysis of a 2012 Ebola virus disease
  epidemic in DRC},
\newblock \emph{Biomath}, \textbf{8} (2019).

\bibitem{DiLauro2022nonMarkov}
\newblock F.~Di~Lauro, W.~R. KhudaBukhsh, I.~Z. Kiss, E.~Kenah, M.~Jensen and
  G.~A. Rempa{\l}a,
\newblock Dynamic survival analysis for non-markovian epidemic models,
\newblock \emph{Journal of The Royal Society Interface}, \textbf{19} (2022),
  20220124.

\bibitem{DiBenedetto10}
\newblock E.~DiBenedetto,
\newblock \emph{Partial differential equations},
\newblock 2nd edition,
\newblock Cornerstones, Birkh\"{a}user Boston, Ltd., Boston, MA, 2010.

\bibitem{Evans10}
\newblock L.~C. Evans,
\newblock \emph{Partial differential equations}, vol.~19 of Graduate Studies in
  Mathematics,
\newblock 2nd edition,
\newblock American Mathematical Society, Providence, RI, 2010.

\bibitem{Tran2009traits}
\newblock R.~Ferri\`ere and V.~C. Tran,
\newblock Stochastic and deterministic models for age-structured populations
  with genetically variable traits,
\newblock in \emph{C{ANUM} 2008}, vol.~27 of ESAIM Proc.,
\newblock EDP Sci., Les Ulis, 2009,
\newblock 289--310.

\bibitem{CDC21}
\newblock C.~for Disease~Control and P.~(CDC),
\newblock {US Centers for Disease Control and Prevention: COVID-19 vaccinations
  in the United States, County},
\newblock
  \urlprefix\url{{https://data.cdc.gov/Vaccinations/COVID-19-Vaccinations-in-the-United-States-County/8xkx-amqh}},
\newblock Accessed: September 8th, 2021.

\bibitem{CDC22}
\newblock C.~for Disease~Control and P.~(CDC),
\newblock {US Centers for Disease Control and Prevention: SARS-CoV-2 Variant
  Classifications and Definitions},
\newblock
  \urlprefix\url{{https://www.cdc.gov/coronavirus/2019-ncov/variants/variant-classifications.html}},
\newblock Accessed: July 13, 2022.

\bibitem{Fournier2004microscopic}
\newblock N.~Fournier and S.~M\'{e}l\'{e}ard,
\newblock A microscopic probabilistic description of a locally regulated
  population and macroscopic approximations,
\newblock \emph{The Annals of Applied Probability}, \textbf{14} (2004),
  1880--1919.

\bibitem{Franco2021Renewal}
\newblock E.~Franco, M.~Gyllenberg and O.~Diekmann,
\newblock One dimensional reduction of a renewal equation for a measure-valued
  function of time describing population dynamics,
\newblock \emph{Acta Applicandae Mathematicae}, \textbf{175} (2021), 12.

\bibitem{Holmdahl20}
\newblock I.~Holmdahl and C.~Buckee,
\newblock Wrong but useful—what covid-19 epidemiologic models can and cannot
  tell us,
\newblock \emph{New England Journal of Medicine}, \textbf{383} (2020),
  303--305.

\bibitem{Hyman2007Infection_age}
\newblock J.~M. Hyman and J.~Li,
\newblock Infection-age structured epidemic models with behavior change or
  treatment,
\newblock \emph{Journal of Biological Dynamics}, \textbf{1} (2007), 109--131.

\bibitem{Jewel20}
\newblock N.~P. Jewell, J.~A. Lewnard and B.~L. Jewell,
\newblock {Predictive Mathematical Models of the COVID-19 Pandemic: Underlying
  Principles and Value of Projections},
\newblock \emph{JAMA}, \textbf{323} (2020), 1893--1894,
\newblock \urlprefix\url{https://doi.org/10.1001/jama.2020.6585}.

\bibitem{kenah2011biostat}
\newblock E.~Kenah,
\newblock {Contact intervals, survival analysis of epidemic data, and
  estimation of $R_0$},
\newblock \emph{Biostatistics}, \textbf{12} (2011), 548--566.

\bibitem{OSU_whitepaper}
\newblock W.~R. KhudaBukhsh, C.~D. Bastian, M.~Wascher, C.~Klaus, S.~Y. Sahai,
  M.~H. Weir, E.~Kenah, E.~Root, J.~H. Tien and G.~A. Rempala,
\newblock {Projecting COVID-19 Cases and Subsequent Hospital Burden in Ohio}, 2022, 
\newblock \emph{medRxiv},
\newblock
  \urlprefix\url{https://www.medrxiv.org/content/10.1101/2022.07.27.22278117v1.full.pdf+html}.

\bibitem{KhudaBukhsh2020InterfaceFocus}
\newblock W.~R. KhudaBukhsh, B.~Choi, E.~Kenah and G.~A. Rempa{\l}a,
\newblock {Survival dynamical systems: individual-level survival analysis from
  population-level epidemic models},
\newblock \emph{Interface Focus}, \textbf{10} (2020).

\bibitem{KhudaBukhsh2020Delay}
\newblock W.~R. KhudaBukhsh, H.-W. Kang, E.~Kenah and G.~Rempa{\l}a,
\newblock Incorporating age and delay into models for biophysical systems,
\newblock \emph{Physical Biology}, \textbf{18} (2021).

\bibitem{KhudaBukhsh2021Prison}
\newblock W.~R. KhudaBukhsh, S.~K. Khalsa, E.~Kenah, G.~A. Rempala and J.~H.
  Tien,
\newblock {COVID-19 dynamics in an Ohio prison}, 2021, 
\newblock \emph{medRxiv},
\newblock
  \urlprefix\url{https://www.medrxiv.org/content/early/2021/01/15/2021.01.14.21249782}.

\bibitem{github}
\newblock C.~Klaus,
\newblock {PDE-DSA github repository}, 2022,
\newblock \urlprefix\url{{https://github.com/klauscj68/PDE-Vax}}.

\bibitem{klaus2022assortative}
\newblock C.~Klaus, M.~Wascher, W.~R. KhudaBukhsh, J.~H. Tien, G.~A. Rempa{\l}a
  and E.~Kenah,
\newblock Assortative mixing among vaccination groups and biased estimation of
  reproduction numbers,
\newblock \emph{The Lancet Infectious Diseases}, \textbf{22} (2022), 579--581.

\bibitem{Koelle22}
\newblock K.~Koelle, M.~A. Martin, R.~Antia, B.~Lopman and N.~E. Dean,
\newblock The changing epidemiology of sars-cov-2,
\newblock \emph{Science}, \textbf{375} (2022), 1116--1121,
\newblock
  \urlprefix\url{https://www.science.org/doi/abs/10.1126/science.abm4915}.

\bibitem{kypraios2017tutorial}
\newblock T.~Kypraios, P.~Neal and D.~Prangle,
\newblock A tutorial introduction to bayesian inference for stochastic epidemic
  models using approximate bayesian computation,
\newblock \emph{Mathematical biosciences}, \textbf{287} (2017), 42--53.

\bibitem{Ladyzhenskaya68}
\newblock O.~A. Lady\v{z}enskaja, V.~A. Solonnikov and N.~N. Ural'ceva,
\newblock \emph{Linear and quasilinear equations of parabolic type},
\newblock Translations of Mathematical Monographs, Vol. 23, American
  Mathematical Society, Providence, R.I., 1968,
\newblock Translated from the Russian by S. Smith.

\bibitem{Driscoll21}
\newblock M.~O'Driscoll, G.~Ribeiro Dos~Santos, L.~Wang, D.~A.~T. Cummings,
  A.~S. Azman, J.~Paireau, A.~Fontanet, S.~Cauchemez and H.~Salje,
\newblock {{A}ge-specific mortality and immunity patterns of
  {S}{A}{R}{S}-{C}o{V}-2},
\newblock \emph{Nature}, \textbf{590} (2021), 140--145.

\bibitem{ODH21}
\newblock O.~D. of~Health,
\newblock {Ohio Department of Health COVID Dashboard},
\newblock
  \urlprefix\url{{https://coronavirus.ohio.gov/wps/portal/gov/covid-19/dashboards/overview}},
\newblock Accessed: October 29, 2021.

\bibitem{Schumaker15}
\newblock L.~L. Schumaker,
\newblock \emph{Spline functions: Computational Methods},
\newblock Society for Industrial and Applied Mathematics, Philadelphia, PA,
  2015,
\newblock {d}oi:10.1137/1.9781611973907.

\bibitem{Sherborne2018nonMarkov}
\newblock N.~Sherborne, J.~C. Miller, K.~B. Blyuss and I.~Z. Kiss,
\newblock Mean-field models for non-markovian epidemics on networks,
\newblock \emph{Journal of Mathematical Biology}, \textbf{76} (2018), 755--778.

\bibitem{ABC}
\newblock S.~A. Sisson, Y.~Fan and M.~A. Beaumont,
\newblock \emph{Handbook of Approximate Bayesian Computation},
\newblock CRC Press, Boca Raton, FL, 2020.

\bibitem{KhudaBukhsh2022Israel}
\newblock I.~Somekh, W.~R. KhudaBukhsh, E.~D. Root, G.~A. Rempa{\l}a,
  E.~Sim{\~o}es and E.~Somekh,
\newblock {Quantifying the Population-Level Effect of the COVID-19 Mass
  Vaccination Campaign in Israel: A Modeling Study},
\newblock \emph{Open Forum Infectious Diseases}, \textbf{9} (2022).

\bibitem{Tran2008limit}
\newblock V.~C. Tran,
\newblock Large population limit and time behaviour of a stochastic particle
  model describing an age-structured population,
\newblock \emph{ESAIM. Probability and Statistics}, \textbf{12} (2008),
  345--386.

\bibitem{vanKampen1998nonMarkov}
\newblock N.~van Kampen,
\newblock {Remarks on Non-Markov Processes},
\newblock \emph{Brazilian Journal of Physics}, \textbf{28} (1998).

\bibitem{Harley2022Ebola}
\newblock H.~Vossler, P.~Akilimali, Y.~Pan, W.~R. KhudaBukhsh, E.~Kenah and
  G.~A. Rempa{\l}a,
\newblock {Analysis of Individual-level Epidemic Data: Study of 2018-2020 Ebola
  Outbreak in Democratic Republic of the Congo},
\newblock \emph{Scientific Reports}, \textbf{12} (2022).

\bibitem{Wascher2021IDSA}
\newblock M.~Wascher, P.~M. Schnell, W.~R. KhudaBukhsh, M.~Quam, J.~H. Tien and
  G.~A. Rempa{\l}a,
\newblock Monitoring sars-cov-2 transmission and prevalence in populations
  under repeated testing, 2021,
\newblock
  \urlprefix\url{https://www.medrxiv.org/content/10.1101/2021.06.22.21259342v1}.

\end{thebibliography}
  
  \providecommand{\href}[2]{#2}
\providecommand{\arxiv}[1]{\href{http://arxiv.org/abs/#1}{arXiv:#1}}
\providecommand{\url}[1]{\texttt{#1}}
\providecommand{\urlprefix}{URL }

\newpage
\appendix
\section{Appendix}
\subsection{Analysis of the coupled flow system \eqref{eq:ys} and \eqref{eq:bdflow_yv}}\label{app:subsecflow}
In this section, we begin by assuming the equation coefficients $\alpha$, $\beta$, $\gamma$, and $\lambda$ are continuous and that $\alpha$, $\beta$, and $\lambda$ also have directional derivatives along characteristics which are continuous as well. We then also assume that the initial data $f_S$ and $f_I$ are continuous, compactly supported, and satisfying the compatibility conditions \eqref{eq:cnstbcs}. As observed in Remark \ref{rmk:app_bdcoeffs} below, we will also come to require that coefficients are bounded.
\subsubsection{Structure conditions}
  First we truncate the domains of the integrals in \eqref{eq:ys} to span their respective $R$'s which we now index from 1 to 3. These $R$'s share a common t-axis but have different s-axes respectively spanning $\left[0,L_i\right]$. Thus, we initially solve a separate problem than \eqref{eq:ys} where instead of integrals over $\left[0,\infty\right)$ we have finite domains. However, because of the compactness of our initial data, this problem will be equivalent to the original formulation \textit{a posteriori}. Letting $y=\left(\ys,\yv,\yi\right)$ and $\mathcal{C}(R)$ stand for the set of continuous functions on the domain $R$, we now rewrite \eqref{eq:ys} in the form 
  \begin{align}
  	\left(\dcl\right)\ys &= -\Lambda_1\left[y\right]\eqcomma \label{eq:pdeclass}\\
  	\left(\dcl\right)\yv &= -\Lambda_2\left[y\right]\eqcomma\nonumber\\
  	\left(\dcl\right)\yi&=-\Lambda_3\left[y\right]\eqcomma \nonumber
  \end{align}
 for functionals $\Lambda_i:\prod^3_{j=1}\mathcal{C}\left(R_j\right)\rightarrow\mathcal{C}\left(R_i\right)$. Moreover, for positive quantities $C_1$ and $\Ld_1$, the $\Lambda_i$ satisfy boundedness and Lipschitz conditions
\begin{align}
	\left\|\Lambda_i\left[y\right]\right\|_\infty &\leq C_1\left(\left\|\lambda\right\|_\infty,\left\|\alpha\right\|_\infty,\left\|\gamma\right\|_\infty,L_I,\left\|y\right\|_\infty\right)\eqcomma \label{eq:Bd_Lambda}\\
	\left\|\Lambda_i\left[y_1\right]-\Lambda_i\left[y_2\right]\right\|_\infty &\leq \Ld_1\left(\left\|\alpha\right\|_\infty,\left\|\beta\right\|_\infty,L_I,\left\|y_1\right\|_\infty,\left\|y_2\right\|_\infty\right)\left\|y_1-y_2\right\|_\infty. \label{eq:Lip_Lambda}
\end{align}
The boundary flow kernels in \eqref{eq:bdflow_yv} are more elaborate but they satisfy similar structure conditions. The key point is that the flow equations are algebraic combinations of functionals which themselves satisfy boundedness and Lipschitz conditions. We may rewrite \eqref{eq:bdflow_yv} as 
\begin{align}
	\partial_t\yv(t,0) = \Gamma_2\left[y\right]\eqcomma\label{eq:bdpdeclass}\\
	\partial_t\yi(t,0) = \Gamma_3\left[y\right]\eqcomma\nonumber
\end{align}
for functionals $\Gamma_i:\prod^3_{j=1}\mathcal{C}\left(R_j\right)\rightarrow\mathcal{C}\left(\left[0,T\right]\right)$. Their boundedness and Lipschitz conditions are
\begin{align}
	\left\|\Gamma_i\left[y\right]\right\|_\infty &\leq C_2\left(\left\|D_\chi\lambda\right\|_\infty,\left\|D_\chi\alpha\right\|_\infty,\left\|D_\chi\beta\right\|_\infty,\left\|\gamma\right\|_\infty,L_S,L_V,L_I,\left\|y\right\|_\infty\right)\eqcomma \label{eq:Bd_Gamma}\\
	\left\|\Gamma_i\left[y_1\right]-\Gamma_i\left[y_2\right]\right\|_\infty &\leq \Ld_2\left(\left\|D_\chi\lambda\right\|_\infty,\left\|D_\chi\alpha\right\|_\infty,\left\|D_\chi\beta\right\|_\infty,\left\|\gamma\right\|_\infty,L_S,L_V,L_I,\left\|y_1\right\|_\infty,\left\|y_2\right\|_\infty\right)\left\|y_1-y_2\right\|_\infty.\label{eq:Lip_Gamma}
\end{align}
Above we let $\left\|D_\chi f\right\|_\infty$ stand for the max supremum norm of both $f$ and $\left(\dcl\right)f$. 
\begin{remark}
	Without loss of generality, we may assume $C_1$, $C_2$, $\Ld_1$, and $\Ld_2$ are increasing functions of their arguments.
\end{remark}
\begin{remark}\label{rmk:app_bdcoeffs}
    The argument dependence of the Lipschitz constants in \eqref{eq:Bd_Lambda}-\eqref{eq:Lip_Lambda} and \eqref{eq:Bd_Gamma}-\eqref{eq:Lip_Gamma} on norms of the equation coefficients quantifies boundedness assumptions needed on equation coefficients for the Lipschitz assumptions to hold. 
\end{remark}
\begin{remark}
	Hereafter, we will abbreviate a constant's dependence on equation coefficients or geometry as \emph{data}. So $C_1=C_1(\data,\left\|y\right\|_\infty)$. Similar holds for $C_2$, $\Ld_1$, and $\Ld_2$.
\end{remark}
\subsubsection{A fixed point characterization of \eqref{eq:pdeclass} and \eqref{eq:bdpdeclass}}
Mimicking the standard approach in \acp{ODE}, we now recast the solution of \eqref{eq:pdeclass} and \eqref{eq:bdpdeclass} as a fixed point for an integral equation. To that end, let $\tau^*=\min(s,t)$ and define \\
$\mathcal{F}_T:\prod^3_{j=1}\mathcal{C}\left(R_j\right)\rightarrow \prod^3_{j=1}\mathcal{C}\left(R_j\right)$ by 
\begin{equation}
\mathcal{F}_T\left[y\right]\left(t,s\right)
=
\begin{cases}
	f_S(s-\tau^*) + \int^{\tau^*}_0\Lambda_1\left[y\right]\big(t-\tau^*+\chi,s-\tau^*+\chi\big) d\chi\\
	\int^{\left(t-\tau^*\right)}_0 \Gamma_2\left[y\right]\left(\theta\right)d\theta + \int^{\tau^*}_0\Lambda_2\left[y\right]\big(t-\tau^*+\chi,s-\tau^*+\chi\big) d\chi\\
	f_I(s-\tau^*) + \int^{\left(t-\tau^*\right)}_0 \Gamma_3\left[y\right]\left(\theta\right)d\theta + \int^{\tau^*}_0\Lambda_3\left[y\right]\big(t-\tau^*+\chi,s-\tau^*+\chi\big) d\chi
\end{cases}\label{eq:fixpt}.
\end{equation}
 Then a fixed point of $\mathcal{F}_T$ is a classical solution of \eqref{eq:ys} and \eqref{eq:bdflow_yv} with the initial conditions matching the prescribed explicit boundary data.
\begin{lemma}
	Let $\mathcal{F}_h:\prod^3_{j=1}\mathcal{C}\left(\bar{R}_j\right)\rightarrow \prod^3_{j=1}\mathcal{C}\left(\bar{R}_j\right)$ where $\bar{R}_i$ has the same s-axis as $R_i$ but in the t-axis only spans $\left[0,h\right]$. Then $\exists M,c,h$ all strictly positive and $c<1$ which depend only on the data so that
	\begin{enumerate}
		\item The restriction $\mathcal{F}_h:\bar{B}_M(0)\rightarrow\bar{B}_M(0)$ and so is bounded on a Banach space,
		\item For $y_1,y_2\in\bar{B}_M(0)$, $\left\|\mathcal{F}_h(y_1)-\mathcal{F}_h(y_2)\right\|_\infty\leq c\left\|y_1-y_2\right\|_\infty$, and so $\mathcal{F}_h$ is a contraction mapping.
	\end{enumerate}
\end{lemma}
\emph{Proof}: For the first part, begin by picking an $M$ larger than twice the supremum norm of $f_S$ and $f_I$ which we label $\left\|f\right\|_\infty$ for brevity. From the sup bounds \eqref{eq:Bd_Lambda} and \eqref{eq:Bd_Gamma}, we see that on $\bar{B}_M(0)$ the vector functionals $\Lambda,\Gamma$ are bounded in terms of this constant M. We may now pick $h$ so that $h\max(\left\|\Lambda\right\|_\infty,\left\|\Gamma\right\|_\infty)<\left\|f\right\|_\infty/3$. Now examining the sums in \eqref{eq:fixpt}, we see these can be no more than $\frac{5}{3}\left\|f\right\|_\infty<M$ owing to the integral domains having length $O(h)$.

For the second part, we may now use that $y_1,y_2$ are supremum bound by $M$ to reduce the dependence of the Lipschitz constants in \eqref{eq:Lip_Lambda} and \eqref{eq:Lip_Gamma} to just the data. But now we conclude in the usual manner by seeing that the integral domains in \eqref{eq:fixpt} have length $O(h)$ and so, by further reduction in $h$, we make the product of $h$ with those Lipschitz constants suitably underneath a threshold $c<1$. \hfill \bbox

\begin{corollary}
	For initial data $\left(f_S,0,f_I\right)$ and an $M$ strictly larger than the supremum norm of the $f$'s, then over a time span $h=h(\data,M)$, the coupled flow equations \eqref{eq:ys} and \eqref{eq:bdflow_yv} have a unique bounded solution in the space $\prod^3_{j=1}\mathcal{C}\left(\bar{R}_j\right)$. This solution may be further taken to be bounded by M over the time span $h$. 
\end{corollary}
\emph{Proof}: This follows from the last lemma and the uniqueness of fixed points for contraction mappings in Banach space. \hfill \bbox
\begin{remark}
	When instead the initial data has a nonzero density $f_V$ for the vaccinated compartment, the same arguments and conclusions hold just with $M$ now also larger than that component's supremum norm.
\end{remark}
\begin{remark}
	Since the size of $h$ is in a monotone decreasing relationship with M but otherwise depends on data which are fixed, we see that so long as an \textit{a priori} global bound can be derived for solutions of \eqref{eq:ys} and \eqref{eq:bdflow_yv}, then they will persist globally and be unique. This follows by the usual iteration of extending the solution by patches $\left[kh,(k+1)h\right]$ along the t-axis.
\end{remark}
\subsubsection{On the equivalence of \eqref{eq:ys} and \eqref{eq:bdflow_yv} with the original formulation \eqref{eq:ys} and \eqref{eq:bdys}-\eqref{eq:bdyi}}
It remains to show that fixed points of \eqref{eq:fixpt} coincide at the t-axis with their intended implicit boundary values \eqref{eq:bdys}-\eqref{eq:bdyi}. The technical concern is that while \eqref{eq:fixpt} is sufficient to imply a fixed point solution is differentiable along characteristics, it does not imply \textit{a priori} that a solution is differentiable along either the t-direction or s-direction individually. Recall in motivating \eqref{eq:bdflow_yv}, we treated the $\partial_s$ and $\partial_t$ derivatives separately. Nevertheless, we show here fixed point solutions do take the intended implicit boundary data. For further examples on the uses of difference quotients and summation by parts in the theory of \acp{PDE}, we refer to standard references such as \cite{Ladyzhenskaya68}.

For the fixed point solution $y$ of \eqref{eq:fixpt}, Let us notate $\phi_V(t) = \int^{L_S}_0 \lambda(t,s)\ys(t,s)ds$. We aim to show $\phi_V(t)=y_V(t,0)$. We will also notate $\Delta_{v}f(t,s) = f(t+v_1,s+v_1)$, which is the finite difference of the function $f$ along the vector $v$. We may now compute
\begin{align*}
    \Delta_{he_1}\phi_V(t) &= \int^{L_S}_0\bigg(\Delta_{he_1}\lambda(t,s)\ys(t+h,s)+\lambda(t,s)\Delta_{he_1}\ys(t,s)\bigg)ds\\
    &= \int^{L_S}_0\bigg(\Delta_{he_1}\lambda(t,s)\ys(t+h,s)+\lambda(t,s)\Delta_{he_1+he_2}\ys(t,s) - \lambda(t,s)\Delta_{he_2}\ys(t+h,s)\bigg)ds.
\end{align*}
We next apply summation by parts to the last term:
\begin{align*}
    -\int^{L_S}_0\lambda(t,s)\Delta_{he_2}\ys(t+h,s)ds &= -\int^{L_S}_0\lambda(t,s)\ys(t+h,s+h)ds +  \int^{L_S}_0\lambda(t,s)\ys(t+h,s)ds\\
    &=-\int^{L_S}_h\lambda(t,s-h)\ys(t+h,s)ds + \int^{L_S}_0\lambda(t,s)\ys(t+h,s)ds\\
    &= -\int^{L_S}_h \Delta_{-he_2}\lambda(t,s)\ys(t+h,s)ds + \int^h_0\lambda(t,s)\ys(t+h,s)ds.
\end{align*}
If we now combine these equations, divide by $h$, and then pass to the limit as $h\rightarrow 0$, we obtain from the regularity of all terms involved that
\begin{align*}
    \partial_t\phi_V(t) &= \int^{L_S}_0\bigg(\ys(t,s)\left(\dcl\right)\lambda(t,s) + \lambda(t,s)\left(\dcl\right)\ys\bigg)ds + \lambda(t,0)\ys(t,0)\\
    &= \int^{L_S}_0\bigg(\left(\dcl\right)\lambda(t,s)\ys(t,s) + \lambda(t,s)\left(-\lambda(t,s)-\int^{L_I}_0\beta(u)\yi(t,u)du\right)\ys(t,s)\bigg)ds.
\end{align*}
Notice in the last equality that we have used directional derivatives of the solution along characteristics are classical and given by \eqref{eq:ys}. Also, we used that $\ys(t,0)=0$ by \eqref{eq:bdys}. Upon inspection, we see this is exactly the flow equation for the boundary data of $\yv$ in \eqref{eq:bdflow_yv}. It now follows from the uniqueness of solutions for \acp{ODE} that $\phi_V(t)$ and $\yv(t,0)$ must coincide. These arguments may be repeated for the $\yi$ solution to show its intended implicit boundary data also evolves according to the flow of \eqref{eq:bdflow_yv}.
\newpage
\subsection{Demonstrating numerical convergence of the implicit boundary conditions}\label{appsubsec:bdconv}
\begin{figure}[h!]
    \centering
    \includegraphics[scale=0.9]{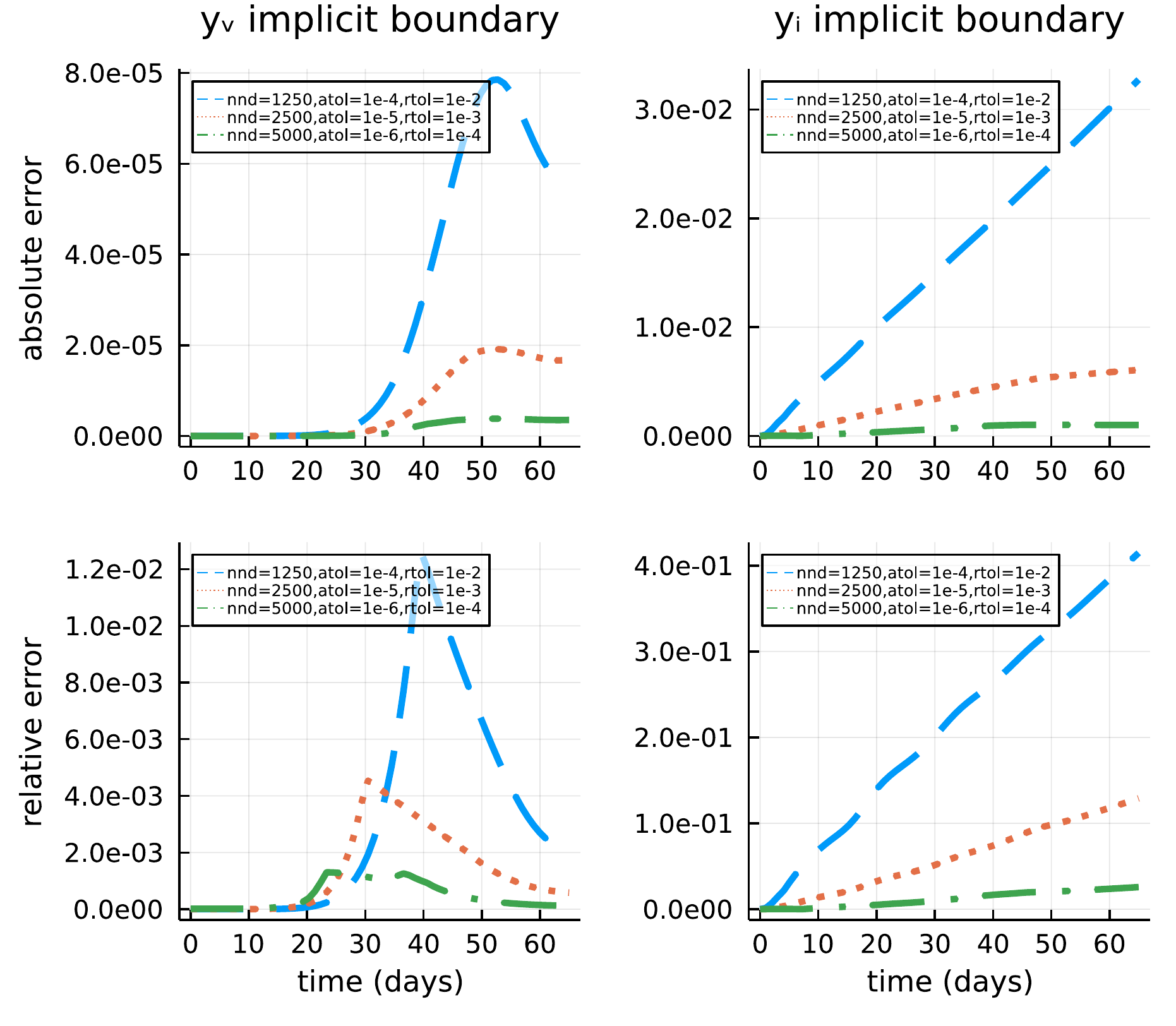}
    \caption{Convergence demonstrations of the PDE \ac{DSA} solver at the implicit boundary axis as the mesh resolution and ode solver tolerances become increasingly fine when using the best fit parameters obtained by \ac{ABC} in Fig~\ref{fig:abcvar} and Table~\ref{tab:prms} of the main body. \textit{(Top)} The absolute error, scaled by the domain length, between the solution's value evaluated at the boundary where implicit data is prescribed and the implicit integral quantity which it is supposed to equal. These are respectively the left and right hand sides of \eqref{eq:bdyv}-\eqref{eq:bdyi}. The left panel is for $y_V$, and the right panel is for $y_I$. \textit{(Bottom)} Similar as the above panels but for the relative error as measured by the magnitude error divided by the solution value at that boundary point.}
    \label{fig:mst_bdconv}
\end{figure}

In Fig~\ref{fig:mst_bdconv} we demonstrate the \ac{PDE} numerical solution better satisfying the implicit boundary conditions as the mesh resolution and ode integrator tolerances become increasingly fine. For the figure legend, we note that nnd is the number of mesh nodes along the s-axis while atol and rtol are respectively the adaptive Euler schemes absolute and relative tolerances for refining the step.  
Recall that, since the magnitude of a probability density varies inversely with the length of its domain, the absolute errors shown are for the solution density multiplied by the size of its domain. In this sense, the absolute error reported is for the solution density if it were first rescaled to a density on the unit interval.

We observe a slight early increase in the relative error of $y_V$'s implicit boundary data as resolutions become increasingly fine. We expect this is due to the true magnitude of $y_V$ becoming increasingly small, which thus leads to division by a smaller number. 
\newpage
\subsection{Parameter posteriors}
\begin{figure}[h!]
    \centering
    \includegraphics[scale=0.9]{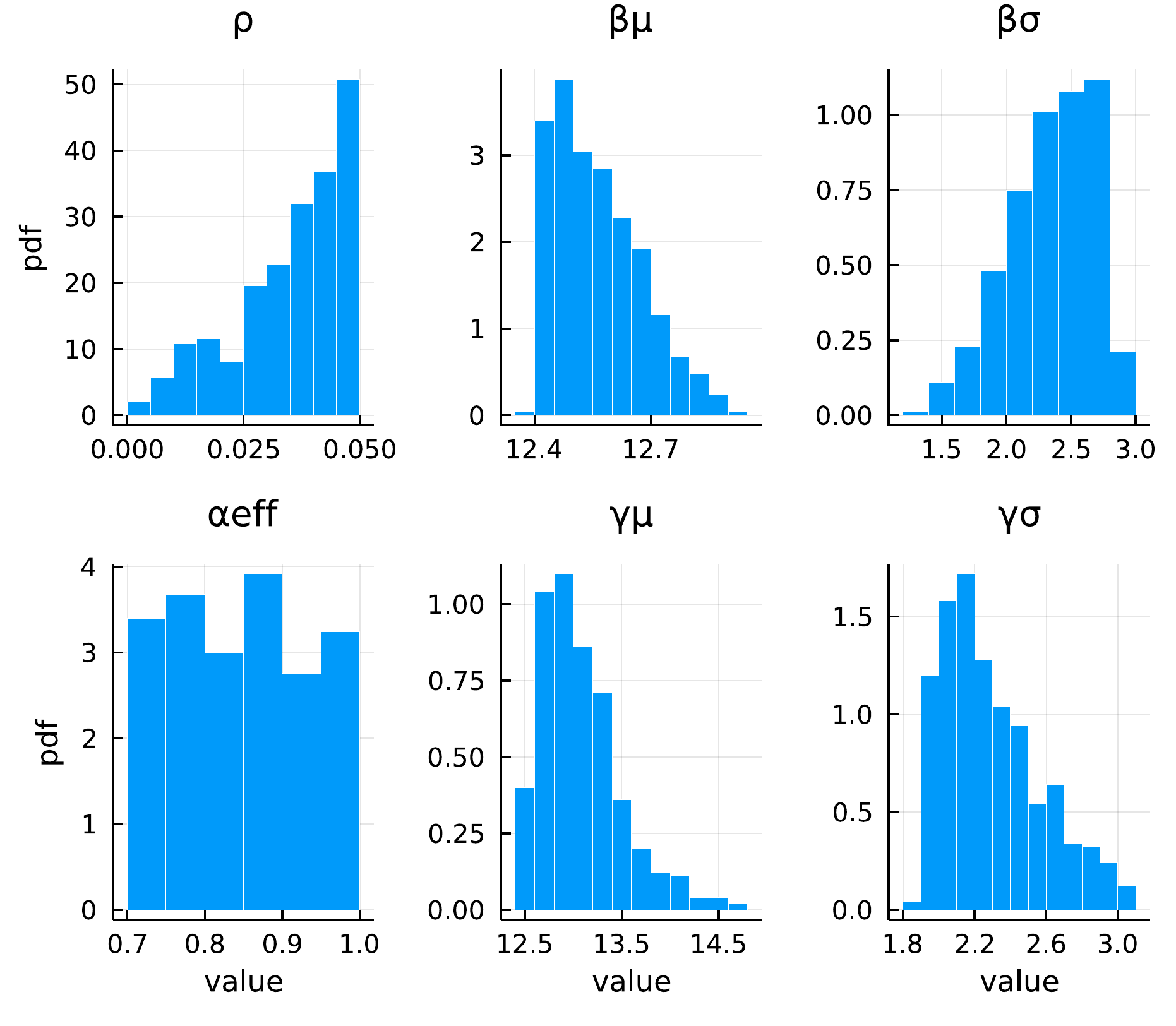}
    \caption{Posterior distributions for \ac{ABC} estimated model parameters. The best 10\% of all 5000 \ac{ABC} samples were kept. The parameter descriptions are given in Table~\ref{tab:prms} of the main body; however, we give the means, $\mu$, and standard deviations, $\sigma$, for the contact interval and infectious period rather than their shape and scale parameters. This amounts to an equivalent parameterization of the two-parameter Weibull distributions.}
    \label{fig:postprm}
\end{figure}

\end{document}